# Control and Characterization of Individual Grains and Grain Boundaries in Graphene Grown by Chemical Vapor Deposition

Qingkai Yu[1,2,*,$], Luis A. Jauregui[3,$], Wei Wu[1,#], Robert Colby[4,#], Jifa Tian[5,#], Zhihua Su[6], Helin Cao[5], Zhihong Liu[6], Deepak Pandey[5], Dongguang Wei[7], sTing Fung Chung[5], Peng Peng[1], Nathan P. Guisinger[8], Eric A. Stach[4,9], Jiming Bao[6], Shin-shem Pei[1], Yong P. Chen[10, *]

[1] Center for Advanced Materials and Department of Electrical and Computer Engineering, University of Houston, Houston, TX 77204

[2] Ingram School of Engineering and Materials Science, Engineering and Commercialization Program, Texas State University, San Marcos, TX 78666

[3] Birck Nanotechnology Center and School of Electrical and Computer Engineering, Purdue University, West Lafayette, IN 47907

[4] Birck Nanotechnology Center and School of Materials Engineering, Purdue University, West Lafayette, IN 47907

[5] Birck Nanotechnology Center and Department of Physics, Purdue University, West Lafayette, IN 47907

[6] Department of Electrical and Computer Engineering, University of Houston, Houston, TX 77204

[7] Carl Zeiss SMT, Inc., One Corporation Way, Peabody, MA 01960

[8] Argonne National Laboratory, Argonne IL 60439

[9] Center for Functional Nanomaterials, Brookhaven National Laboratory, Upton NY 11973

[10] Birck Nanotechnology Center and Department of Physics and School of Electrical and Computer Engineering, Purdue University, West Lafayette, IN 47907

[$] These authors contributed equally.     [#] These authors contributed equally.
[*] To whom correspondence should be addressed: qingkai.yu@txstate.edu; yongchen@purdue.edu

**Abstract**

The strong interest in graphene has motivated the scalable production of high quality graphene and graphene devices. Since large-scale graphene films synthesized to date are typically polycrystalline, it is important to characterize and control grain boundaries, generally believed to degrade graphene quality. Here we study single-crystal graphene grains synthesized by ambient CVD on polycrystalline Cu, and show how individual boundaries between coalescing grains affect graphene's electronic properties. The graphene grains show no definite epitaxial relationship with the Cu substrate, and can cross Cu grain boundaries. The edges of these grains are found to be predominantly parallel to zigzag directions. We show that grain boundaries give a significant Raman "D" peak, impede electrical transport, and induce prominent weak localization indicative of intervalley scattering in graphene. Finally, we demonstrate an approach using pre-patterned growth seeds to control graphene nucleation, opening a route towards scalable fabrication of single-crystal graphene devices without grain boundaries.

The extraordinary properties and potential applications of graphene[1-3] have motivated the development of large-scale, synthetic graphene grown by various methods, such as graphitization of SiC surfaces[4,5] and chemical vapor deposition (CVD) on transition metals such as Ni[6-8] and Cu[9]. In particular, it has been shown that large and predominantly monolayer graphene of excellent quality can be synthesized by CVD on polycrystalline Cu foils[9-11]. This relatively simple and low-cost method has been used to produce graphene that can reach impressive sizes (e.g. 30 inches, the largest graphene ever made by any method)[10] and can be easily transferred to other substrates[10, 12]. However, the large-scale synthetic graphene films produced so far are typically polycrystalline[13, 14], consisting of many single-crystalline grains separated by grain boundaries[15-17]. In the growth of such polycrystalline graphene, graphene grains nucleate from random and uncontrolled locations. As the growth of such grains proceeds, they coalesce and eventually form an interconnected polycrystalline film. The grain boundaries (which, by definition, are defective) are expected to degrade the electrical[13,14] and mechanical[18] properties of the resulting films. It is well known that the availability of high quality, large single-crystal Si wafers is foundational to the present Si-based electronics[19]. In order for graphene to realize its promise in "carbon-based" electronics, it will clearly be necessary to synthesize either large-scale, high-quality single-crystalline graphene films, or to achieve better control over the nucleation of individual graphene grains and to avoid the grain boundaries in fabricated graphene devices. In this article, we study graphene grains (either isolated grains or a small number of merged grains) formed during the early stage of ambient CVD growth on Cu foils. We obtain fundamental insight into the growth mechanisms of single-crystalline graphene on polycrystalline Cu substrates. The hexagonally-shaped grains (with sizes of tens of microns) have their edges macroscopically oriented predominantly parallel to zigzag directions. This display of particular orientations can facilitate studies of crystal direction dependent phenomena in graphene. We also report measurements of how individual grain boundaries impede electronic transport and scatter charge carriers. Finally we demonstrate controlled graphene nucleation and synthesis of graphene grain arrays that could enable scalable fabrication of devices free of detrimental grain boundaries.

The graphene studied in this work was synthesized on polycrystalline Cu foils by ambient CVD (see Method), using procedures largely similar to those described in our previous publications[11,20]. We halted growth before the graphene grains merged with each other to form a globally continuous (but polycrystalline) graphene film[11,20]. Fig. 1a is a typical optical microscopy image of the Cu surface after CVD growth, showing many graphene islands, which consist of either a single grain or a few coalesced grains. Fig. 1b is a scanning electron microscope (SEM) image of several graphene grains grown within one single Cu grain. The graphene grains were typically hexagonally-shaped, with ~120° corners, suggesting that their edges are parallel to specific crystallographic directions. The hexagonal shape is notably different from the flower-like shape of previously reported graphene grains grown by low pressure CVD [9,13,21,22] (notably, we also obtain similar flower-shaped graphene grains using our CVD setup operated under low pressure, see Supplemental Fig. S1. A detailed study of the dependence of the shape of graphene grains on various growth conditions is beyond the scope of this work and will be presented elsewhere). All of the graphene grains shown in Fig. 1b appear to have orientations (edge directions) closely aligned with each other, which would suggest some well defined epitaxial relation between the graphene lattice and that of underlying Cu grain. However, a closer examination of the image reveals that the alignment is in fact not precise. Furthermore, we have also frequently observed situations where the individual graphene grains grown within a single Cu grain have very different orientations with each other, as shown in Fig. 1c. The lack of correlation between the crystal orientation of the graphene and the underlying Cu indicates that the interaction between the graphene and Cu is weak, and that there is no definite epitaxial relationship between the two. Recent experimental (STM)[23] and theoretical (van der Waals-density functional calculation)[24] studies of graphene on single-crystal Cu (111) have also found a very weak graphene-Cu interaction. Interestingly, we have found that individual graphene grains can be grown continuously (without any apparent distortion of its hexagonal shape) across Cu grain boundaries, as shown in Fig 1d (see also Supplemental Fig. S2). This phenomenon reflects the weak influence of the Cu crystal lattice on graphene growth and demonstrates that single-crystalline graphene can grow on polycrystalline Cu.

Establishing the crystal structure and orientation of the graphene grains is relatively straightforward using transmission electron microscopy (TEM). The graphene grains used in the measurements were transferred to ~100 nm thick amorphous SiN membranes. Fig. 2a shows a TEM image of a hexagonally shaped graphene grain and its characteristic selected area electron diffraction (SAED) pattern (inset). Only one set of six-fold symmetric diffraction spots was observed, indicating that the grain is a single crystal. There was occasionally a small degree of arc to the diffraction spots collected from larger graphene grains. This can be attributed to cracks, folds, tears or residues due to the transfer process. One can also determine the crystal direction of a grain's edges by comparing the orientation of the grain in real space with the orientation of the diffraction pattern (as shown schematically in Supplemental Fig. S3). We found that most of the grain edges (despite their microscopic roughness) were approximately aligned with zigzag directions (indicated by dashed lines in the example of Fig. 2a), while armchair directions were seen only very rarely. Sometimes the edges were partially folded or torn, as also seen in Fig. 2a, a feature not uncommon in our samples and likely a result of the transfer process (notably, this also limits our ability to obtain atomic-resolution TEM images of the edge structure). Additional examples of graphene grains and their edge orientations (as determined by TEM/SAED) can be found in Supplemental Fig. S5. Fig. 2b presents a TEM image of two coalescing graphene grains with Fig. 2c-d showing their respective SAED patterns. While each grain is seen to be a single crystal, the crystal orientations of the two are rotated from each other (by ~28° in this case).

Fig. 3 presents scanning tunneling microscopy (STM) images taken from a representative graphene grain grown on Cu. A large-scan-area STM topography image of the grain near a corner is shown in Fig. 3a. The angle between the two edges (indicated by the dashed lines) is ~120°. At this length scale, we found that the graphene surface showed significant roughness and height fluctuation (on the order of 10 nm, see Supplemental Fig. S6). Figs. 3b-3d show atomically-resolved STM topography images (filtered to improve contrast) taken from 3 different locations (marked by different color squares in Fig. 3a) in the graphene grain. The sample orientation and the scanning orientation were kept same as used in Fig. 3a. The characteristic honeycomb lattice (highlighted with a few model hexagons superimposed on the images) of single layer graphene can be clearly observed. The 3 images show the same lattice orientation (within our experimental uncertainty related to a small tip motion hysteresis and sample/tip drift), consistent with the single-crystalline nature of the grain. Furthermore, the two edges in Fig. 3a are seen to each be parallel to a zigzag direction in the graphene lattice (Figs. 3b-3d, where we have indicated two zigzag ("Z") and one armchair ("A") directions with correspondingly-labeled arrows). Our STM results thus confirm that the graphene grain studied is a single layer, and that it is single-crystalline with edges along zigzag directions.

The relative stability of two types of edges along major crystallographic directions in graphene (zigzag versus armchair) has been a question under active investigation both experimentally[25-28] and theoretically[29-31]. For example, exfoliated monolayer graphene flakes can show both zigzag and armchair oriented edges[1,32]. On the other hand, edges created by certain etching reactions[27], or in holes formed by electron beam irradiation[25] in graphene, appear to favor zigzag directions. Previously studied graphene "nanograins" (with sizes much smaller than those studied here) that were epitaxially grown on various single-crystal metal surfaces[33-35] show mostly zigzag edges, while those grown on SiC (0001) show armchair edges[28]. Such a rich array of behavior suggests that the relative stability of zigzag versus armchair edges may be strongly influenced by graphene's environment (such as the substrate). Our findings indicate that zigzag edges are also preferred in graphene grown on polycrystalline Cu (even when definite epitaxial relations may not exist). Furthermore, since our relatively large sized graphene grains (spanning tens of microns) have macroscopically zigzag-oriented edges (serving as intrinsic "direction markers"), such grains may facilitate the study of crystal direction dependent physics in graphene[36-38] and the fabrication of graphene nanostructures with well-defined edge orientations.

Raman spectroscopy is a powerful technique to identify the number of graphene layers and the presence of defects in graphene[39-41]. We have performed Raman spectroscopy and mapping with a 532 nm excitation laser on CVD grown graphene grains transferred onto $SiO_2$/Si wafers (see Methods). The intensities ($I_x$, where x = D, G or 2D) of characteristic graphene Raman peaks[39], D (~1350 cm$^{-1}$), G (~1580 cm$^{-1}$) and 2D (~2690 cm$^{-1}$) were extracted and their spatial dependences (Raman maps) are plotted in Figs. 4(a – c) for a single graphene grain, and in Figs. 4(d - f) for two coalesced graphene grains, respectively. Several representative Raman spectra are also shown in Supplemental Fig. S8. The typical $I_{2D}$ is more than twice that of $I_G$, indicating that our samples are single layer graphene [9,11,39]. Additionally, $I_D$ (see also 3D plots in Supplemental Fig. S9) is negligibly small (indicating a low defect content[39-41]) over most of the area within each graphene grain, with the notable exception of a few isolated spots displaying relatively large $I_D$ (e.g. location "c" in Fig. S8, indicating defects). We have mapped many (>20) graphene grains and found that every grain contains at least one such defect spot (with large $I_D$). We suggest some of these defect sites could be nucleation centers, i.e. where the growth is initiated. A pronounced $I_D$ was also observed on the edges of the grains, consistent with previous Raman studies of graphene edges [39-41], as well as at the grain boundary between two coalesced grains (Figs. 4d and S8). Raman mapping of the D peak intensity provides a particularly convenient way to clearly identify the locations of grain boundaries (which are otherwise more difficult to visualize, for example, in SEM images shown in Fig. 1, and atomic force microscope (AFM) image shown in Supplemental Fig. S10). The Raman D peak reflects an (elastic) intervalley scattering process[39, 40]. Our observation of large $I_D$ at the grain boundaries suggests that they are a significant source of intervalley scattering, which gives prominent signatures (weak localization) in electronic transport measurements to be presented next.

Electrical transport measurements were performed on multi-terminal devices fabricated from graphene grains transferred onto SiO$_2$/Si wafers (Methods). Fig. 5a shows a representative device ("a") fabricated on two coalesced graphene grains that meet at a single grain boundary. Multiple electrodes were patterned to contact each grain to allow simultaneous measurements of both intra-grain (within the grain) and inter-grain (across the grain boundary) transport. Fig. 5b shows representative intra-grain and inter-grain current-voltage (I-V) curves measured at room temperature for this device. All the I-V curves are linear (ohmic). The resistances extracted from the slopes of these I-V curves are $R_L$ ~550Ω (left grain), $R_R$ ~200Ω (right grain) and $R_{CG}$~3kΩ (across the grain boundary), respectively. Taking into account of the device geometry, we can extract from $R_{CG}$ an effective inter-grain resistivity (based on a geometric average of the sample width, see Fig. S11) of $\rho_{CG} \sim 5k\Omega/\square$. This is higher than both of the intra-grain resistivities ($\rho_L \sim 2000\Omega/\square$ for the left grain and $\rho_R \sim 400\Omega/\square$ for the right grain, extracted from the corresponding resistances above), reflecting the effect of grain boundary to impede the electrical transport. We can also calculate an inter-grain series resistance (r) that neglects the grain boundary, by integrating the intra-grain resistivities (see Fig. S11). This gives r~ 0.9 k$\Omega$, much smaller than the measured $R_{CG}$ (~3k$\Omega$), indicating that the grain boundary provides an "extra" resistance (~2.1 k$\Omega$ in this case). We have found qualitatively similar results in all of the coalesced grain devices we have measured (Supplemental Fig. S12): inter-grain resistivities ($\rho_{CG}$) are always higher than corresponding intra-grain resistivities ($\rho_G$, on each side of the grain boundary), and inter-grain resistances ($R_{CG}$) are always higher than the calculated inter-grain series resistances (r).

Magnetotransport measurements were performed for device "a" under a perpendicular magnetic field (B). Fig. 5c presents low temperature (4.3K) magnetoresistance ($R_{xx}(B)$) measurements across the grain boundary compared to $R_{xx}(B)$ measured within each of the graphene grains. The inter-grain $R_{xx}(B)$ displays a prominent peak at B = 0 T, associated with weak localization (WL). Such a WL peak was much weaker or even unobservable for intra-grain $R_{xx}(B)$. Similar results were found in multiple devices, and an example for another device ("b") measured at an even lower temperature of 450 mK is shown in Supplemental Fig. S13. WL results from interplay between impurity scattering and quantum coherent transport of carriers (such interplay also leads to reproducible "universal conductance fluctuations" (UCF) in the resistance[42], which were also observable in our devices). Raising the temperature (*T*) is expected to destroy the phase coherence, and thus diminish the WL feature, as was indeed observed (Fig. 5d, showing temperature dependence of the inter-grain $R_{xx}(B)$). The inter-grain WL feature we observed at low temperature can be well fitted to the WL theory developed for graphene[43], allowing us to extract various inelastic (phase-breaking) and elastic (intervalley and intravalley) scattering lengths (Supplemental Fig. S14). In graphene, due to the chiral nature of carriers[43-45], WL requires the presence of sharp lattice defects that cause intervalley scattering with large momentum ($q$) transfer. Our observation of prominent inter-grain WL but much weaker or negligible intra-grain WL indicates that grain boundaries are major sources of intervalley scattering in our graphene devices (where inter-grain current has to cross the grain boundary), while such scattering is less significant within the single-crystal graphene grains. This is consistent with the relatively large value of intervalley scattering length $L_i$ (~200 nm) extracted from the low $T$ weak localization fitting (Fig. S14), as a carrier would have to encounter a grain boundary, a graphene edge, or some other isolated and more localized lattice defects within the grain in order to undergo intervalley scattering.

The intra-grain mobility ($\mu_G$) extracted from low temperature Hall measurements in all of the devices we studied (fabricated on isolated as well as coalesced grains) ranged from <$10^3$ cm$^2$/Vs to ~$10^4$ cm$^2$/Vs. The devices were all found to be p-type doped (likely due to adsorbates from the environment and residues from the transfer and device fabrication processes). Inter-grain mobility was not extracted as the two coalescing grains often had different carrier densities (in the cases where their densities were comparable, the inter-grain mobility was lower than $\mu_G$ by similar factors as the resistivity ratios shown in Fig. S12a). Our work has clearly indicated the detrimental effect of grain boundaries on electronic transport, and that avoiding grain boundaries is beneficial for improving the mobility. However, the wide variation of $\mu_G$ in different samples (sometimes even between neighboring grains) and the occasional low $\mu_G$ observed suggest that other sources of disorder could also strongly affect the mobilities. Such "extra" disorder may also be partially responsible for the small intravalley scattering length ($L_* < 20$ nm, Fig. S14) observed (indicating a significant amount of small-$q$ scattering defects such as charged impurities, line defects and ripples etc[2, 43, 45] that may arise from, e.g. graphene transfer or device fabrication processes[13]). Improving related fabrication processes to reduce such defects will be required to achieve consistent high mobilities in graphene-based devices.

Due to the detrimental effect of grain boundaries on graphene's electronic properties and device performance, it is desirable to avoid grain boundaries when fabricating graphene devices. This would be very difficult to achieve for a large number of graphene devices in a practical circuit if the graphene grains nucleate at random locations, as shown in Fig. 1. Here we demonstrate a method to control the nucleation of graphene grains using seeded growth, and to synthesize spatially ordered arrays of graphene grains with pre-determined locations (Fig. 6). Seeded growth is commonly used in the growth of single-crystal materials (for example the well-known Czochralski process used to create single-crystal Si[19]). Unlike the previously discussed growth of randomly nucleated graphene grains (Fig. 1), here seed crystals were placed on Cu as nucleation centers. Fig. 6a shows an example array of seed crystals lithographically patterned from a continuous multilayer graphene film pre-grown by CVD on a Cu foil[46, 47]. Following patterning, the Cu foil was re-inserted into the CVD furnace to perform re-growth. Fig. 6b and 6c show typical results after seeded growth with a shorter and longer growth time, respectively, giving rise to smaller and larger average grain sizes. Monolayer graphene (confirmed by Raman spectroscopy) grains have been successfully synthesized from seed crystals made of either multilayer or monolayer CVD graphene as well as transferred exfoliated graphene/graphite. Seeds made from multilayer CVD graphene have mostly been chosen thus far, being both easy to pattern into large area seed arrays and sufficiently robust to withstand the patterning process. From Figs. 6b and 6c, it is apparent that the seeded growth resulted in largely ordered arrays of graphene grains, with each grain growing from a seed crystal (many of the seeds were still visible inside the synthesized grains shown in Fig. 6b). In Fig. 6d, we compare seeded grains with randomly nucleated grains grown concurrently on the same Cu foil (which has a pre-patterned seed array to the left of the dashed line, and no seeds to the right). The seeded growth led to mostly ordered arrays of grains of a much higher density than the randomly nucleated grains. In our seeded growth demonstrated so far, randomly nucleated (i.e. non-seeded) grains were also occasionally (but more rarely) observed (e.g. in Fig. 6d, as well as a particular example in Fig. 6b marked by the arrow at the lower left). Occasionally, multiple (merged) grains, apparently growing from a single shared seed, were also observed. These observations may reflect the competition between seeded growth and randomly-nucleated growth, and possible effects of the seeds we used. More work is underway to determine whether and how different types and properties of seeds may affect the seeded growth, and to further improve the yield of seeded single-crystal graphene grains. Compared to the regular CVD growth of graphene based on random nucleation studied earlier, the seeded growth demonstrated here could offer a viable strategy to control the nucleation of graphene crystals, making an array of many graphene grains with pre-determined locations. Such an addressable graphene grain array could facilitate large scale fabrication of single crystalline graphene electronic devices that avoid grain boundaries altogether, without the need of starting with a large single-crystalline graphene sheet (then subject to subsequent etching and patterning).

In conclusion, we have synthesized hexagonally shaped graphene single-crystal grains of up to tens of microns in size on polycrystalline Cu. The single-crystal nature of these graphene grains is confirmed by TEM and STM, and their edges are found to be predominantly parallel to zigzag directions. Individual grain boundaries are characterized and found to cause weak localization and impede electrical transport. We have also demonstrated that single-crystal graphene growth is not limited by the polycrystallinity of the Cu substrate and can be artificially initiated by a seed, paving the way for controllable synthesis of single-crystal graphene and large scale fabrication of single-crystalline graphene devices free of grain boundaries.

**Methods**

*Graphene synthesis and transfer.* Graphene grains were grown by CVD ($CH_4$ as the carbon feedstock) on Cu substrates at ambient pressure. First, a Cu foil (25-μm-thick, 99.8%, Alfa Aesar) was loaded into a CVD furnace and heated up to 1050 ºC under 300 sccm Ar and 10 sccm $H_2$. After reaching 1050 ºC, the sample was annealed for 30 min or longer without changing the gas flow rates. The growth was then carried out at 1050 ºC for ~10 min under a gas mixture of 300 sccm diluted (in Ar) $CH_4$ (concentration 8 ppm) and 10 sccm of $H_2$. Finally, the sample was rapidly cooled to room temperature under the protection of Ar and $H_2$, then taken out of the furnace for characterizations. Graphene samples were transferred by a PMMA (polymethyl methacrylate) assisted process in a Cu etchant (iron nitrate) onto ~100 nm thick amorphous $Si_xN_y$ membranes (chosen for their flat surface over large areas) for TEM studies, or onto to $SiO_2$/Si wafers (doped Si covered by 300 nm $SiO_2$) for Raman and electrical transport studies.

*Transmission electron microscopy (TEM)*. All TEM images were taken at 80 kV on an FEI Tecnai 20. Selected area electron diffraction (SAED) patterns and bright field images were taken in plan view, with the beam at or near parallel to 〈0001〉 for the graphene. Montages are spliced together from individual bright field images using a 2D stitching plugin (http://fly.mpi-cbg.de/~preibisch/software.html) with Fiji[48]. To elucidate contrast between the graphene reflections and background (from amorphous SiN) in diffraction patterns, a mild low pass filter was sometimes applied (>3 pixels), followed by a locally adaptive histogram equalization filter (50 pixel blocksize, 256 histogram bins, max. slope 8.0)[48, 49]. Nonetheless, graphene diffraction spots are quite clear in the original, unfiltered patterns when viewed at full size (see Supplemental Fig. S4). SAED patterns examined from multiple areas in each grain, and when possible from a larger area including the entire grain, demonstrate that the grains studied are primarily single-crystalline graphene.

*Scanning tunneling microscopy (STM)*. STM measurements were carried out in an Omicron ultra-high vacuum (UHV, base pressure < $10^{-11}$ mbar) scanning tunneling microscope at room temperature (300 K). Prior to the measurements, the sample was annealed in UHV at 300°C for 12 hours to remove the surface adsorbates and contaminants. The chemically etched STM tip was made of tungsten or platinum/iridium alloy and was also annealed before imaging. The atomic resolution images have been processed with a wavelet-based filter[50] to enhance the contrast.

*Raman Spectroscopy and Mapping*. Raman spectroscopy and spatial Raman mapping were performed using a Horiba Jobin Yvon Xplora confocal Raman microscope equipped with a motorized sample stage from Marzhauser Wetzlar (00-24-427-0000). The wavelength of the excitation laser was 532 nm and the power of the laser was kept below 2 mW without noticeable sample heating. The laser spot size was ~0.6 μm with a 100X objective lens (numerical aperture =0.90). All the Raman maps had a pixel size of 0.4 μm for both x and y directions. The spectral resolution was 2.5 cm$^{-1}$ (using a grating with 1200 grooves/mm) and each spectrum was an average of 3 acquisitions (5 seconds of accumulation time per acquisition). The intensity of a Raman peak was extracted from the maximum value after baseline subtraction over corresponding spectral range (1300-1400 cm$^{-1}$ for "D", 1560-1600 cm$^{-1}$ for "G" and 2620-2760 cm$^{-1}$ for "2D").

*Device fabrication and electronic transport measurements*. The electrical contacts (Cr/Au, 5nm/35nm, e-beam evaporated) to graphene grains were patterned by e-beam lithography. No oxygen plasma etching for patterning graphene was performed to avoid introducing extra defects in graphene. The resistances (Figs. 5c and 5d) were measured using low frequency lock-in detection (SR 830) with a driving current of 1μA. The I-V curves in Fig. 5b were measured with a DC source meter (Keithley 2400). All the electrical transport data in the main text were taken in vacuum (~10$^{-5}$ torr) in a variable temperature probe station (Lakeshore CPX-VF).

**References**

1. Geim, A. K. & Novoselov, K. S. The rise of graphene. *Nature Mater.* **6,** 183-191 (2007).
2. Castro Neto, A. H., Guinea, F., Peres, N. M. R., Novoselov, K. S., & Geim, A. K. The electronic properties of graphene. *Rev. Mod. Phys.* **81,** 109-162 (2009).
3. Geim, A. K. Graphene: Status and prospects. *Science* **324,** 1530-1534 (2009).
4. de Heer, W. A. et al. Epitaxial graphene. *Solid State Commun.* **143,** 92-100 (2007).
5. Emtsev, K. V. et al. Towards wafer-size graphene layers by atmospheric pressure graphitization of silicon carbide. *Nature Mater.* **8,** 203-207 (2009).
6. Yu, Q. K. et al. Graphene segregated on Ni surfaces and transferred to insulators. *Appl. Phys. Lett.* **93,** 113103 (2008).
7. Reina, A. et al. Large area, few-layer graphene films on arbitrary substrates by chemical vapor deposition. *Nano Lett.* **9,** 30-35 (2009).
8. Kim, K. S. et al. Large-scale pattern growth of graphene films for stretchable transparent electrodes. *Nature* **457,** 706-710 (2009).

9. Li, X. S. et al. Large-area synthesis of high-quality and uniform graphene films on copper foils. *Science* **324,** 1312-1314 (2009).

10. Bae, S. et al. Roll-to-roll production of 30-inch graphene films for transparent electrodes. *Nature Nanotech.* **5,** 574-578 (2010).

11. Cao, H. L. et al. Electronic transport in chemical vapor deposited graphene synthesized on Cu: quantum Hall effect and weak localization. *Appl. Phys. Lett.* **96,** 122106 (2010).

12. Chen, Y. P. & Yu, Q. K. Nanomaterials: Graphene rolls off the press. *Nature Nanotech.* **5,** 559-560 (2010).

13. Li, X. S. et al. Graphene films with large domain size by a two-step chemical vapor deposition process. *Nano Lett.* **10**, 4328–4334 (2010).

14. Yazyev, O. V. & Louie, S. G. Electronic transport in polycrystalline graphene. *Nature Mater.* **9,** 806-809 (2010).

15. Huang, P. Y. et al. Imaging grains and grain boundaries in single-layer graphene: An atomic patchwork quilt. *Nature* **469,** 389-392 (2011).

16. Kim, K. et al. Grain Boundary Mapping in Polycrystalline Graphene, *ACS Nano*, Article ASAP (2011). DOI: 10.1021/nn1033423

17. An, J. et al. Domain (grain) boundaries and evidence of twin like structures in CVD grown graphene. Preprint at http://arxiv.org/abs/1010.3905 (2010).

18. Grantab, R., Shenoy, V.B., & Ruoff, R.S. Anomalous strength characteristics of tilt grain boundaries in graphene. *Science* **330**, 946-948 (2010).

19. Plummer, J. D., Deal, M. D., & Griffin, P. B., *Silicon VLSI Technology*. (Prentice Hall, Upper Saddle River, NJ, 2000).

20. Wu, W. et al. Wafer-scale synthesis of graphene by chemical vapor deposition and its application in hydrogen sensing. *Sens. and Actuators B: Chem.* **150,** 296-300 (2010).

21. Li, X. S., Cai, W. W., Colombo, L., & Ruoff, R. S. Evolution of graphene growth on Ni and Cu by carbon isotope labeling. *Nano Lett.* **9,** 4268-4272 (2009).

22. Li, X. et al. Large-Area Graphene Single Crystals Grown by Low-Pressure Chemical Vapor Deposition of Methane on Copper. *J. Am. Chem. Soc.* **133**, 2816-2819 (2011)

23. Gao, L., Guest, J. R., & Guisinger, N. P. Epitaxial graphene on Cu(111). *Nano Lett.* **10,** 3512-3516 (2010).

24. Vanin, M. et al. Graphene on metals: A van der Waals density functional study. *Phys. Rev. B* **81,** 081408 (2010).

25. Girit, C. O. et al. Graphene at the edge: Stability and dynamics. *Science* **323,** 1705-1708 (2009).

26. Jia, X. T. et al. Controlled formation of sharp zigzag and armchair edges in graphitic nanoribbons. *Science* **323,** 1701-1705 (2009).

27. Nemes-Incze, P., Magda, G., Kamaras, K., & Biro, L. P. Crystallographically selective nanopatterning of graphene on $SiO_2$. *Nano Res.* **3,** 110-116 (2010).

28. Rutter, G. M., Guisinger, N. P., Crain, J. N., First, P. N., & Stroscio, J. A. Edge structure of epitaxial graphene islands. *Phys. Rev. B* **81,** 245408 (2010).

29. Okada, S. Energetics of nanoscale graphene ribbons: Edge geometries and electronic structures. *Phys. Rev. B* **77,** 041408 (2008).

30. Nakajima, T. & Shintani, K. Molecular dynamics study of energetics of graphene flakes. *J. Appl. Phys.* **106,** 114305 (2009).

31. Gan, C. K. & Srolovitz, D. J. First-principles study of graphene edge properties and flake shapes. *Phys. Rev. B* **81,** 125445 (2010).

32. Neubeck, S. et al. Direct determination of the crystallographic orientation of graphene edges by atomic resolution imaging. *Appl. Phys. Lett.* **97,** 053110 (2010).

33. Coraux, J. et al. Growth of graphene on Ir(111). *New J. Phys.* **11,** 023006 (2009).

34. Eom, D. et al. Structure and electronic properties of graphene nanoislands on Co(0001). *Nano Lett.* **9,** 2844-2848 (2009).

35. Yamamoto, M., Obata, S., & Saiki, K. Structure and properties of chemically prepared nanographene islands characterized by scanning tunneling microscopy. *Surf. Interface Anal.* **42,** 1637-1641 (2010).

36. Son, Y. W., Cohen, M. L., & Louie, S. G. Half-metallic graphene nanoribbons. *Nature* **444,** 347-349 (2006).

37. Huang, M. Y. et al. Phonon softening and crystallographic orientation of strained graphene studied by Raman spectroscopy. *Proc. Natl. Acad. Sci. USA* **106,** 7304-7308 (2009).

38. Mohiuddin, T. M. G. et al. Uniaxial strain in graphene by Raman spectroscopy: G peak splitting, Gruneisen parameters, and sample orientation. *Phys. Rev. B* **79,** 205433 (2009).

39. Ferrari, A. C. et al. Raman spectrum of graphene and graphene layers. *Phys. Rev. Lett.* **97,** 187401 (2006).

40. Ferrari, A. C. Raman spectroscopy of graphene and graphite: Disorder, electron-phonon coupling, doping and nonadiabatic effects. *Solid State Commun.* **143,** 47-57 (2007).

41. Malard, L. M., Pimenta, M. A., Dresselhaus, G., & Dresselhaus, M. S. Raman spectroscopy in graphene. *Phys. Rep.* **473,** 51-87 (2009).

42. Beenakker, C. W. J. & van Houten, H. *Quantum Transport in Semiconductor Nanostructures* Vol. 44 (Academic Press, New York, 1991).

43. McCann, E. et al. Weak-localization magnetoresistance and valley symmetry in graphene. *Phys. Rev. Lett.* **97,** 146805 (2006).

44. Morozov, S. V. et al. Strong suppression of weak localization in graphene. *Phys. Rev. Lett.* **97,** 016801 (2006).

45. Tikhonenko, F. V., Horsell, D. W., Gorbachev, R. V., & Savchenko, A. K. Weak localization in graphene flakes. *Phys. Rev. Lett.* **100,** 056802 (2008).

46. Bhaviripudi, S., Jia, X. T., Dresselhaus, M. S., & Kong, J. Role of kinetic factors in chemical vapor deposition synthesis of uniform large area graphene using copper catalyst. *Nano Lett.* **10,** 4128-4133 (2010).

47. Lee, S., Lee, K., & Zhong, Z. Wafer scale homogeneous bilayer graphene films by chemical vapor deposition. *Nano Lett.* **10**, 4702–4707 (2010).

48. Available at http://pacific.mpi-cbg.de/wiki/index.php/Fiji.

49. Available at CLAHE plugin, http://pacific.mpi-cbg.de/cgi-bin/gitweb.cgi?p=mpicbg.git;a=tree;f=mpicbg/ij/clahe.

50. Gackenheimer, C., Cayon, L., & Reifenberger, R. Analysis of scanning probe microscope images using wavelets. *Ultramicroscopy* **106,** 389-397 (2006).

**Acknowledgements**

QY acknowledges support from NSF and UHCAM. NPG acknowledges support from DOE SISGR. EAS acknowledges support from DOE BES. JB acknowledges support from TcSUH and the Welch Foundation. YPC acknowledges support from NSF, DTRA, DHS, IBM, Miller Family Endowment and Midwest Institute for Nanoelectronics Discovery (MIND). STM measurements were carried out at Argonne National Laboratory under the support of DOE user program.

**Author contributions**

QY led the synthesis and YPC led the characterization efforts. QY, WW, ZL and SZ synthesized the graphene samples and performed SEM. RC and EAS performed the TEM. JT, HC and NG performed the STM. LAJ performed the Raman measurements. LAJ and HC fabricated the devices and performed electronic transport measurements. QY, LAJ, RC, JT, EAS and YPC wrote the paper and all authors contributed to the discussions.

**Competing financial interests**

The authors declare no competing financial interests.

**Figures**

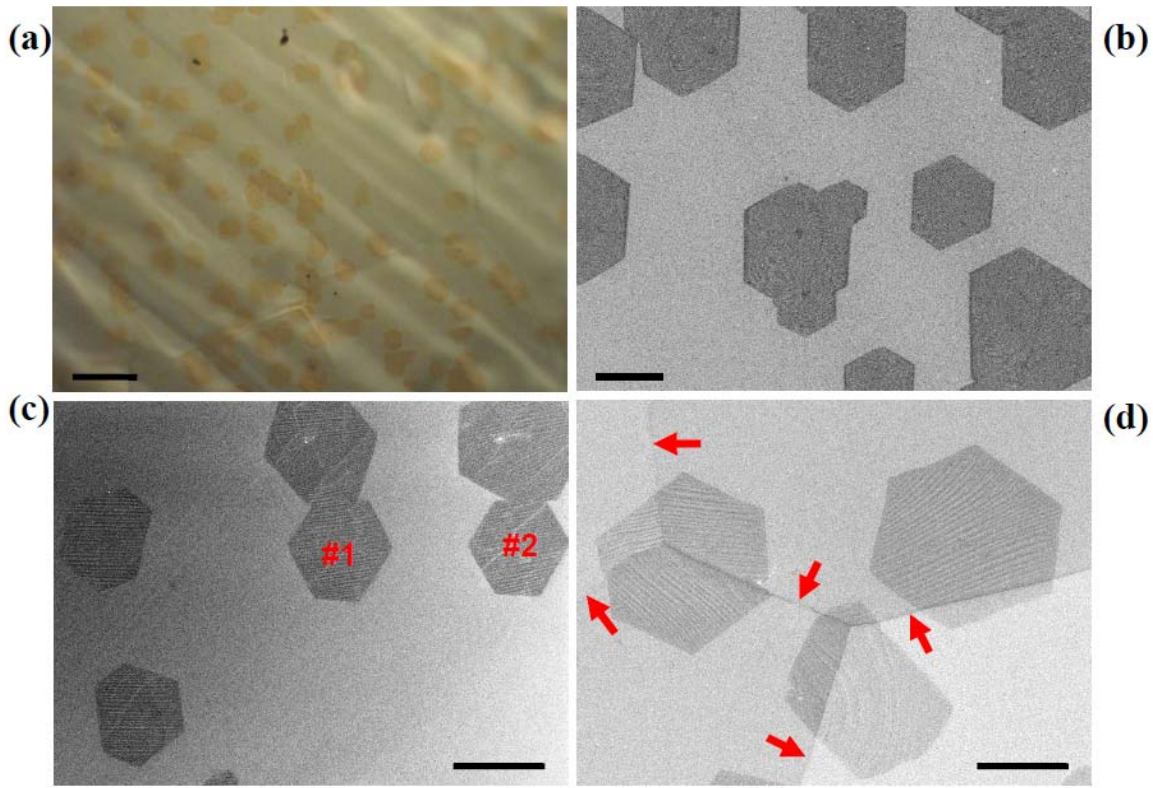

Figure 1. **Graphene grains grown on Cu substrates**. (a) An optical microscopy image of as-grown, mostly hexagonally shaped graphene grains on Cu. Some grains are seen to coalesce to form larger islands. (b) SEM image of as-grown grains whose edge orientations are approximately aligned with each other. (c) SEM image of as-grown grains whose edge orientations are not aligned with each other (except for the two grains labeled as #1 and #2). Images (b) and (c) were each taken from within one Cu crystal grain. (d) SEM image showing that hexagonally-shaped graphene grains can be grown continuously across Cu crystal grain boundaries (indicated by red arrows). The scale bars in (a)-(d) are 25 μm, 10 μm, 10 μm and 5 μm, respectively.

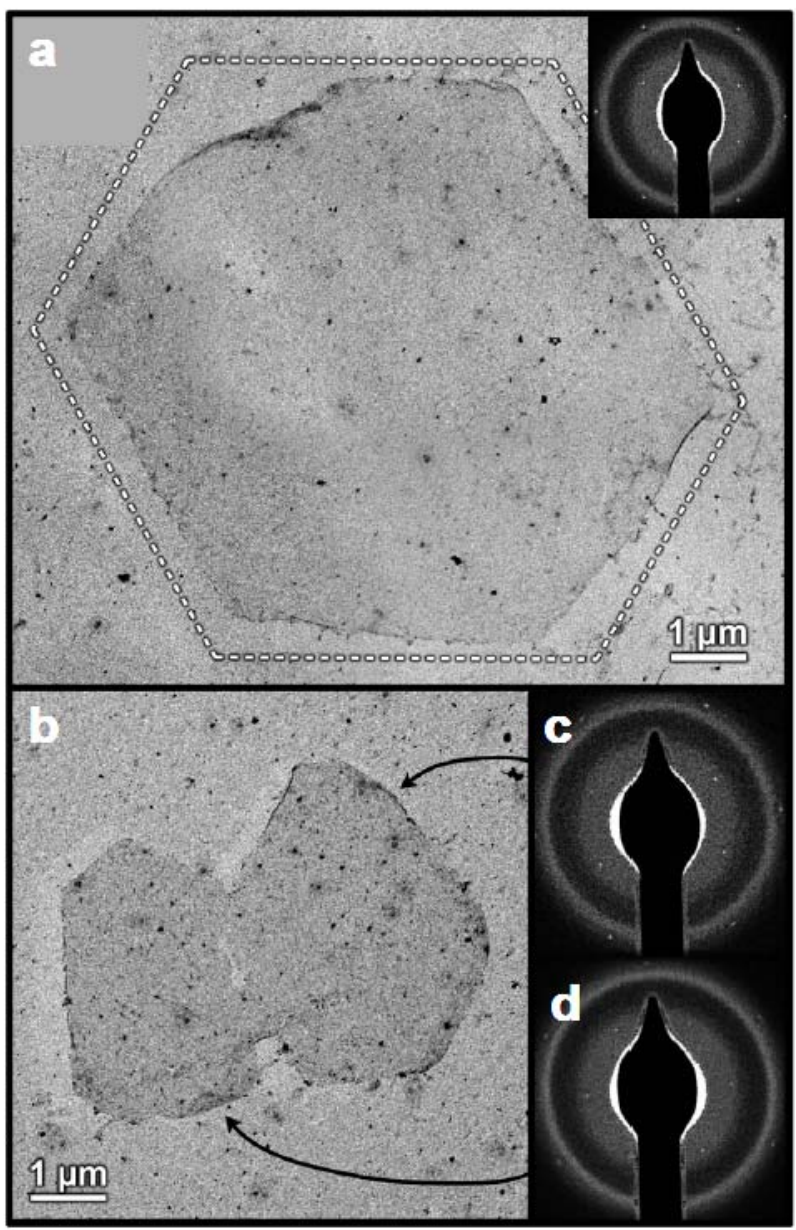

Figure 2. **Transmission electron microscopy (TEM) and electron diffraction from graphene grains**. (a) A montage of bright field TEM images (80 kV) spliced together to show an example of a graphene grain. A schematic outline has been included based on the adjoining SAED pattern (inset) to demonstrate that the edges of the graphene grains are mostly parallel to the zigzag directions (dashed lines). The edges typically curl as a result of the transfer from the growth substrate to a TEM-compatible support. (b) Bright field TEM image of 2 coalesced grains and SAED patterns (c,d) from the individual grains demonstrate that each corresponds to a single crystal of graphene, and that the two

grains are rotated from each other by approximately 28°. The SAED patterns have been filtered (the unfiltered versions are shown in Supplemental Fig. S4) to improve the contrast over the diffuse background, contributed mostly by the amorphous SiN support. The dark pointer in each pattern is the shadow of the beam stop used to block the intense direct beam.

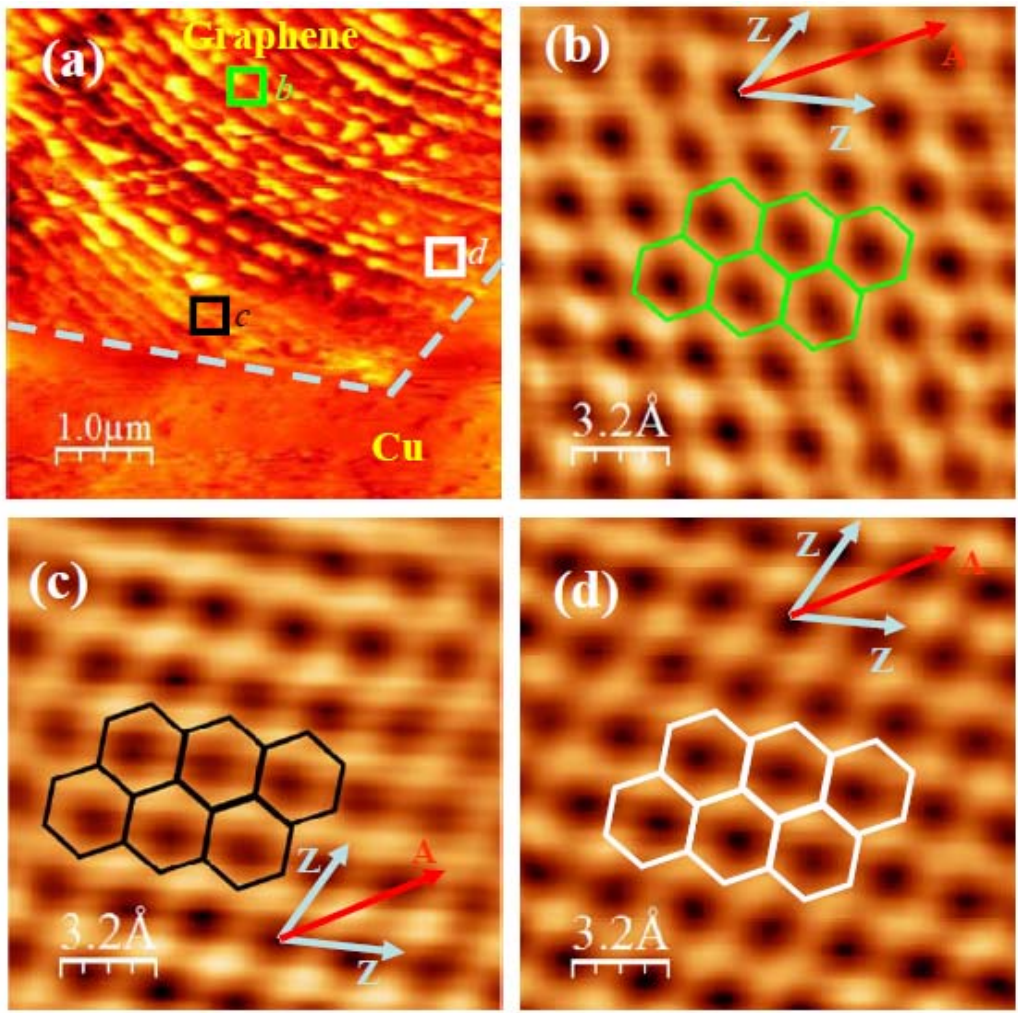

Figure 3. **Scanning tunneling microscopy (STM) of a single-crystal graphene grain on Cu**: (a) STM topography image taken near a corner of a graphene grain on Cu. The image was acquired with sample–tip bias $V_b = -2$ V and tunneling current $I=50$ pA. Dashed lines mark the edges of this grain. (b-d) Atomic-resolution STM topography images (filtered to improve contrast) taken from 3 different areas in the grain as indicated in (a). The green (b), black (c) and white (d) squares (not to scale) indicate the approximate locations where the images were taken ($V_b=-0.2$ V, $I=20$ nA). A few model hexagons are superimposed on the images to demonstrate the graphene honeycomb lattice. Select special crystal directions ("Z" for zigzag, "A" for armchair) are indicated by arrows. A small distortion in images (c) and (d) was due to a slight hysteresis in the movement of the STM tip. The corresponding unfiltered raw images of (b-d) are shown in Supplemental Fig. S7.

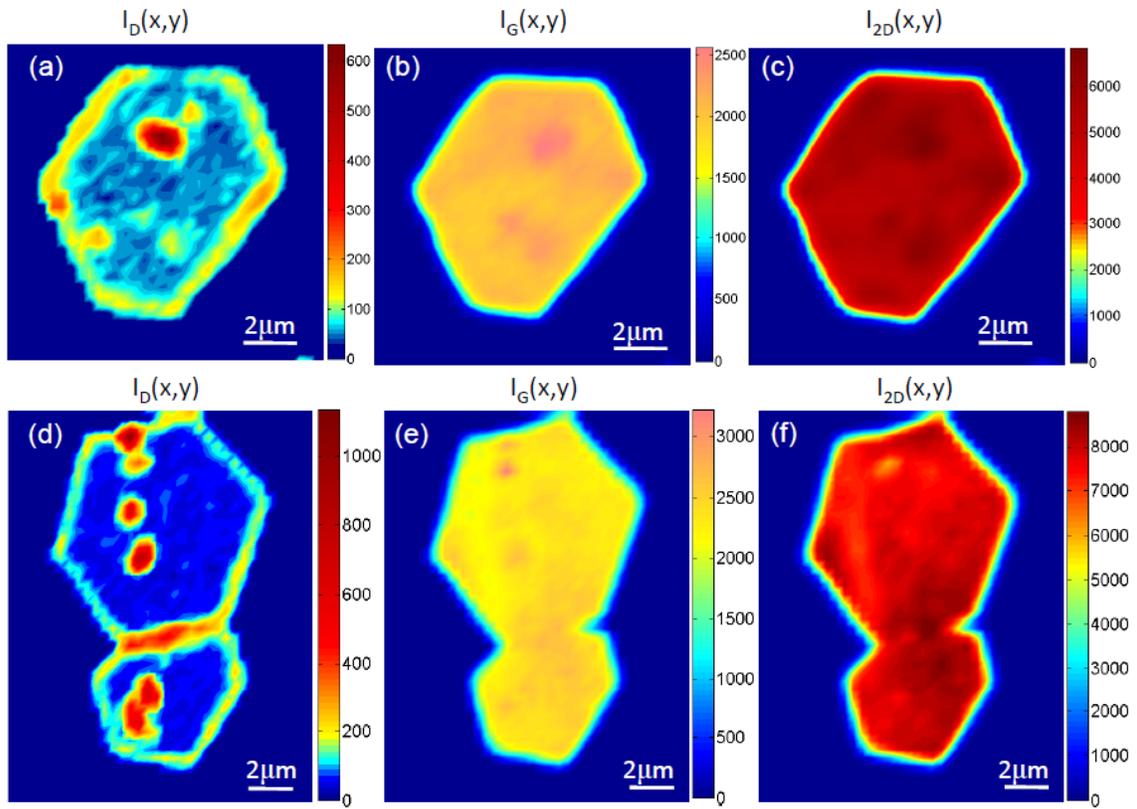

Figure 4. **Spectroscopic Raman mapping of graphene grains and grain boundaries**: (a-c) Intensity maps of the "D", "G" and "2D" bands, respectively, for a single-crystal graphene grain. (d-f) Intensity maps of the "D", "G" and "2D" bands, respectively, for two coalesced graphene grains with a single grain boundary. The wavelength of the Raman excitation laser is 532 nm. The spectral resolution is 2.5 cm$^{-1}$. The Raman map pixel size is 0.4 μm.

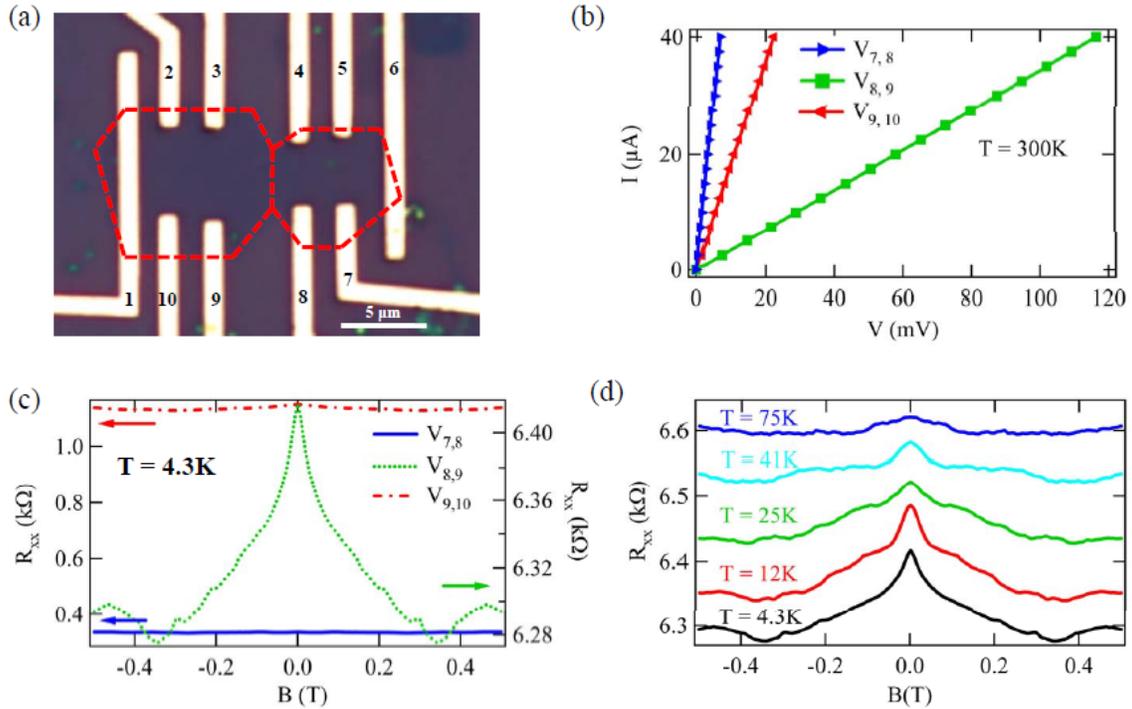

Figure 5. **Electronic transport cross a single grain boundary.** (a) Optical image of a device with multiple electrodes (numbered 1-10) contacting two coalesced graphene grains (indicated by dashed lines). (b) Representative room-temperature I-V curves measured within each graphene grain and across the grain boundary. The measurements shown were performed at zero magnetic field, and using 4-probe configurations, with contacts "1" and "6" as current leads, and 3 pairs of voltage leads labeled in the legend. (c) Four-terminal magnetoresistance ($R_{xx}$) measured at 4.3 K within each graphene grain and across the grain boundary (using the same set of contacts as in b). The inter-grain $R_{xx}$ (dotted curve) displays a prominent weak localization (WL) peak. (d) Temperature dependence of the WL feature (c) in the inter-grain $R_{xx}$. Traces are offset vertically for clarity. $R_{xx}$ data in (c-d) have been symmetrized between two opposite magnetic field directions to remove a small linear background from a mixing of Hall component.

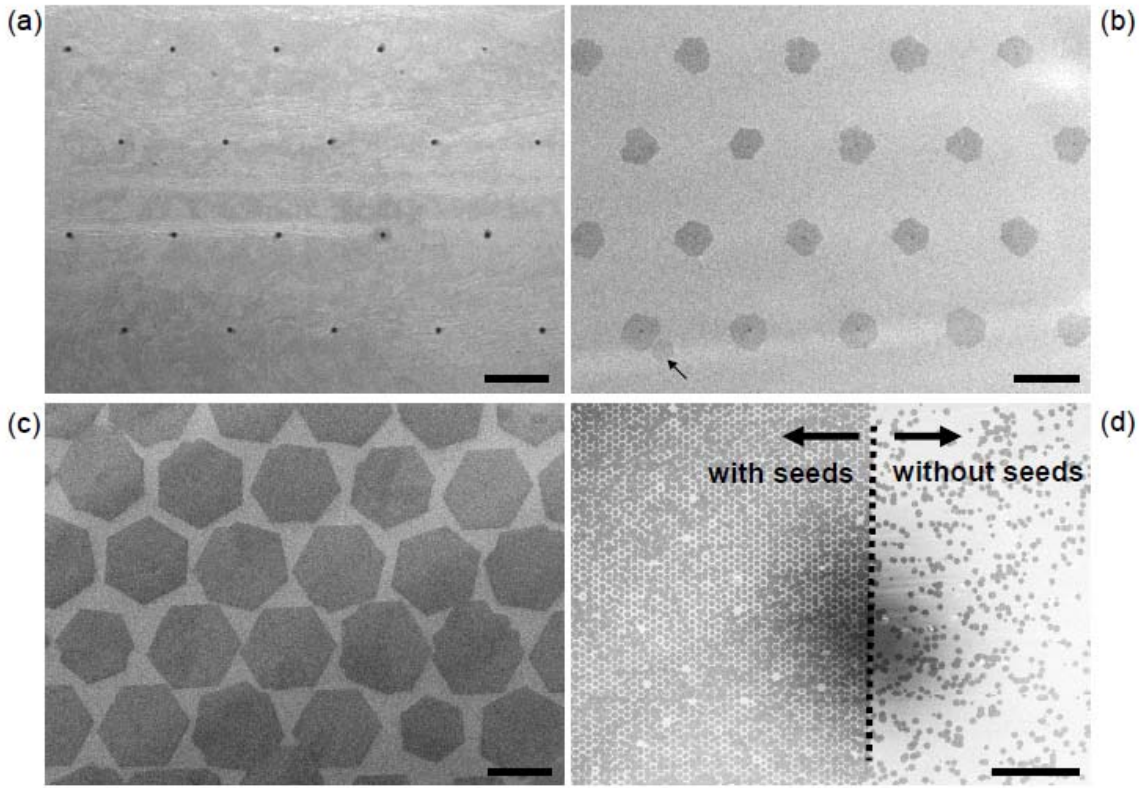

Figure 6. **Seeded growth of graphene grains**. (a) SEM image showing an array of seed crystals (seen as dots) patterned from a pre-grown multilayer graphene film on Cu foil by e-beam lithography. The period of the array is 16 μm. The size of each seed is about 500 nm. (b) SEM image of a typical graphene grain array grown from an array of seed crystals, with a relatively short growth time (5 min). The seeds can be seen at the centers of many grains. A grain that nucleated randomly (i.e. not from one of the pre-patterned seeds) is also observed (indicated by the arrow at the lower left). (c) SEM image of a graphene grain array from seeded growth similar to (b), but following a longer growth time (15 min). The representative images (a-c) do not necessarily correspond to the exactly the same area on the Cu foil. (d) Low magnification SEM image of a seeded array of graphene grains (left to the dotted line), next to a randomly-nucleated set of graphene grains in an area without seeds (right to the dotted line). Scale bars in (a-c) are 10 μm and the scale bar in (d) is 200 μm. To reduce Cu surface defects that may facilitate random (not from seeds) nucleation of graphene grains, the Cu foil was annealed for 3 hours before the seeded growth.

**Supplemental Information**

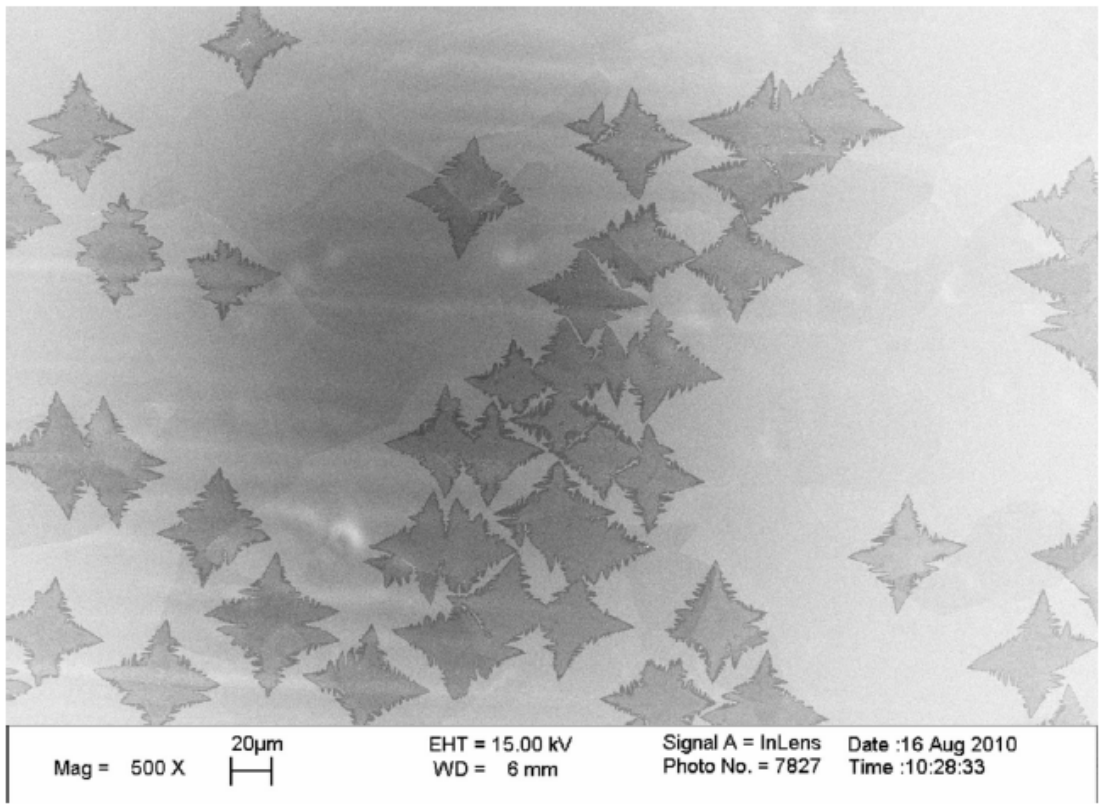

**Figure S1.** SEM image of the graphene grains grown by low pressure CVD (500 ppm CH$_4$ 300 sccm, 945 mTorr at downstream). Such flower-like graphene grains are similar in shape to those grown previously using low pressure CVD [S1-S3], but are very different from the hexagonally-shaped graphene grains shown in the main text grown by ambient CVD.

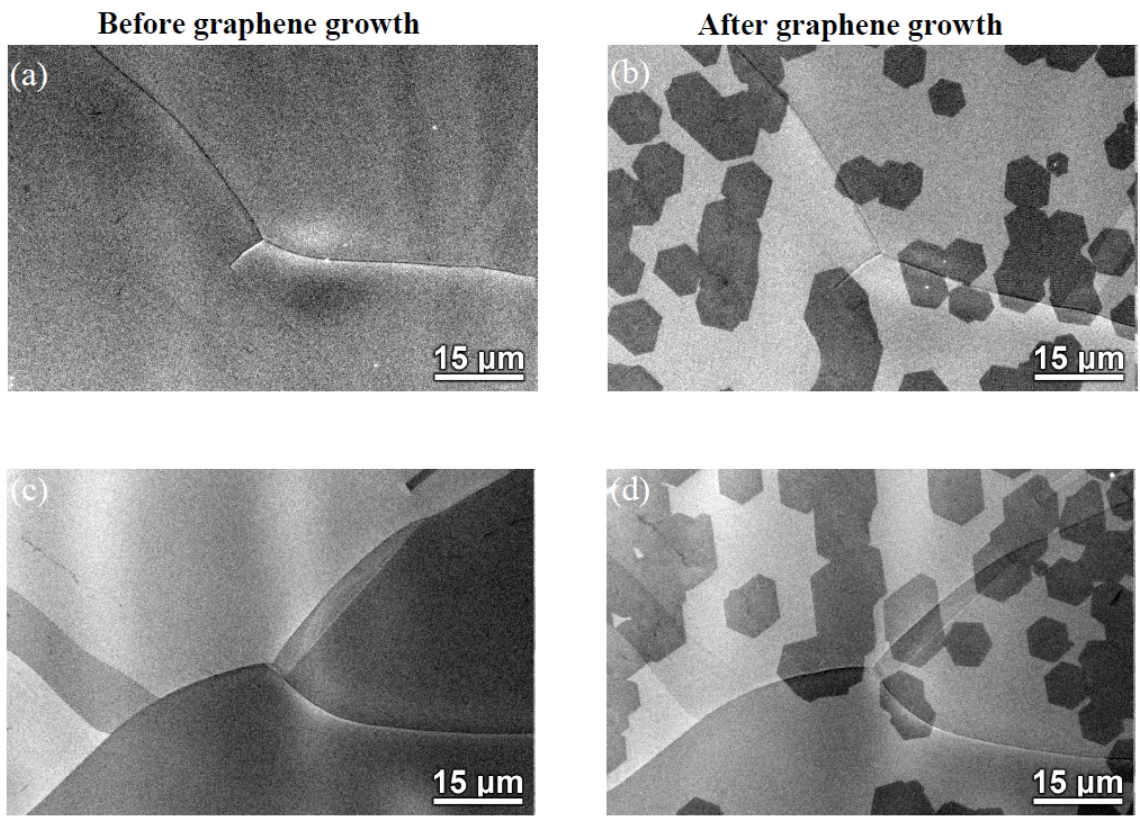

Figure S2. SEM images of an annealed Cu foil taken out of the furnace before (a,c) and after (b,d) graphene growth, respectively. Images (c,d) were taken from a different area of the foil from images (a,b). We observe similar grain structures of the annealed Cu foil before and after the graphene growth, suggesting that the change of Cu grain boundaries is relatively small during our graphene growth.

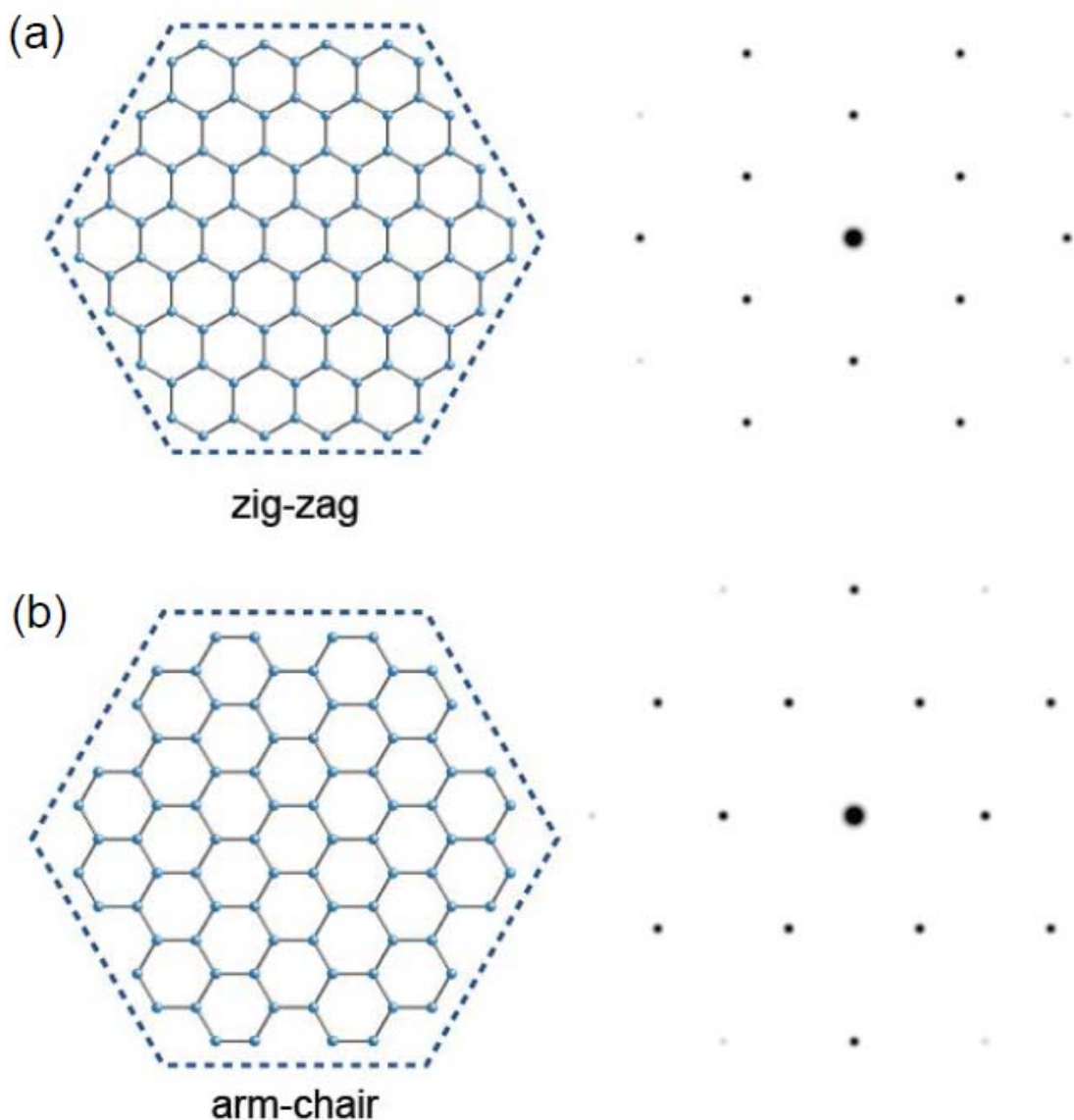

**Figure S3**. Cartoon schematics of graphene single-crystal grains with zigzag (a) and armchair (b) edges, with adjoining calculated parallel beam electron diffraction patterns, created using CrystalMaker®. The orientation of the edges of a grain relative to armchair or zigzag directions can be likewise obtained by comparing TEM images of that grain with a diffraction pattern from the same. The orientation between the TEM images and SAED patterns in our measurements has also been calibrated using a SiC cross-section imaged along $<11\bar{2}0>$.

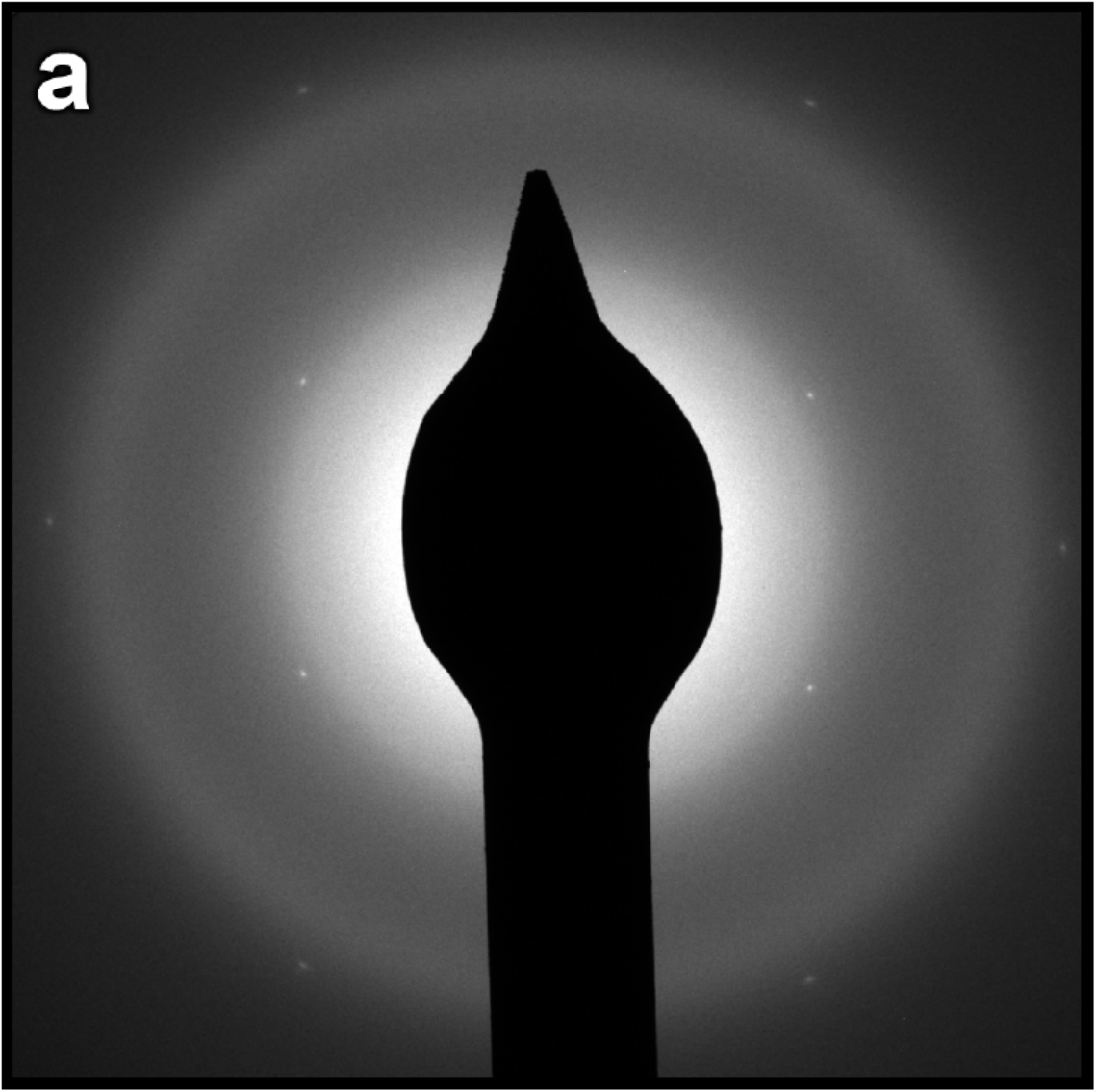

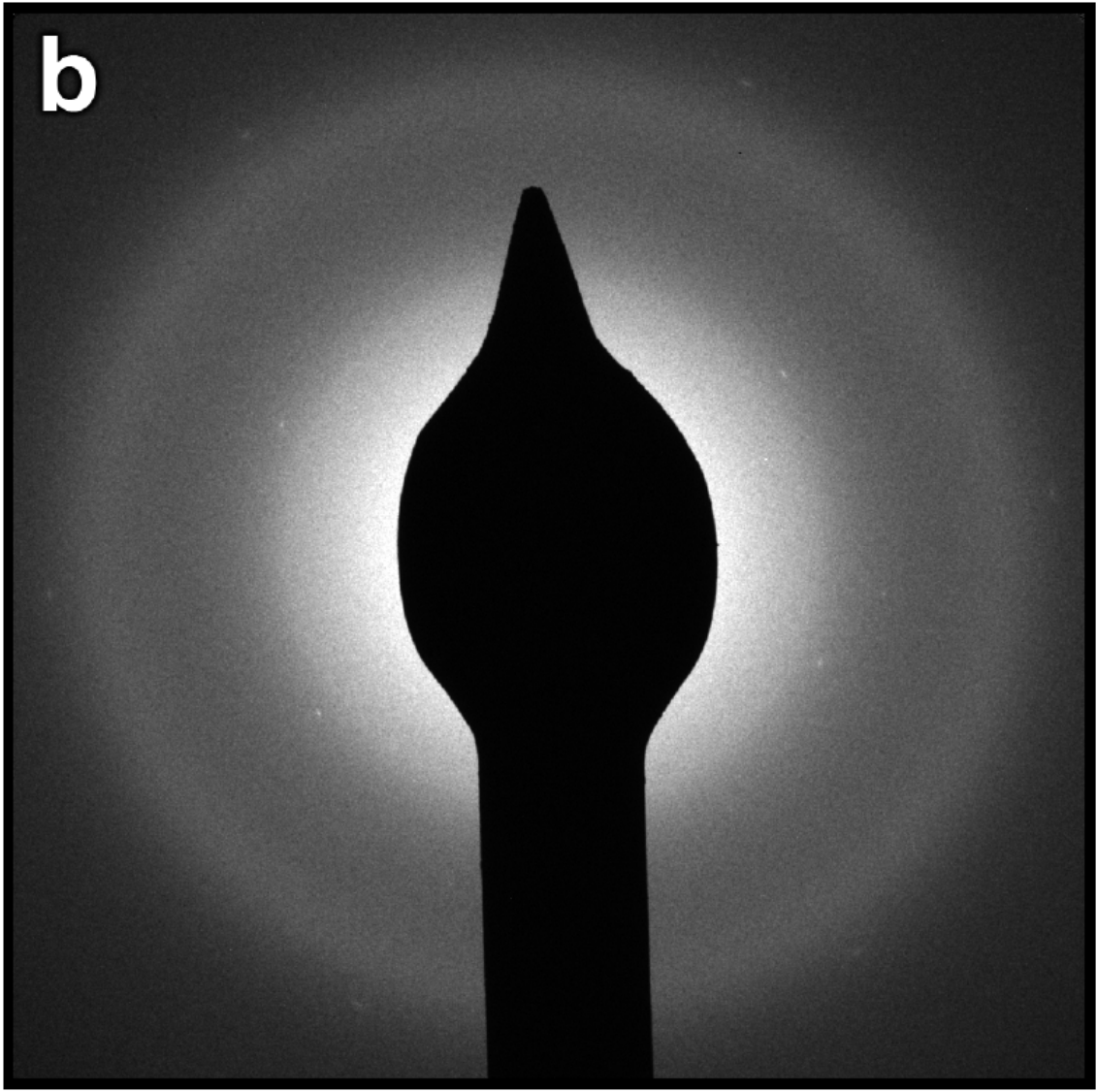

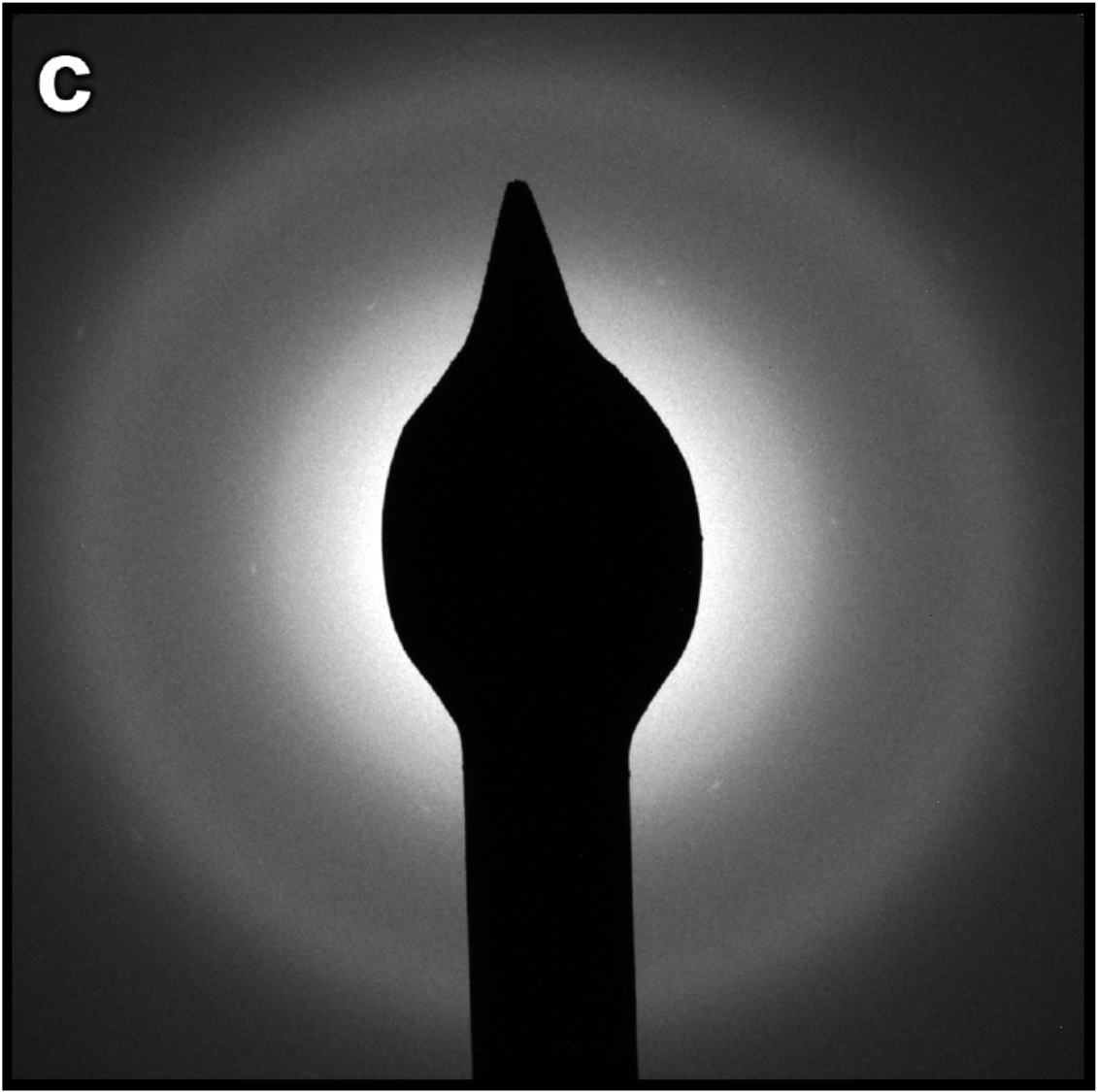

**Figure S4**. Original, unprocessed SAED patterns of the graphene grains corresponding to the processed (filtered) versions shown in main text (Fig. 2): (a) for Fig. 2a inset; (b) for Fig. 2c; (c) for Fig. 2d.

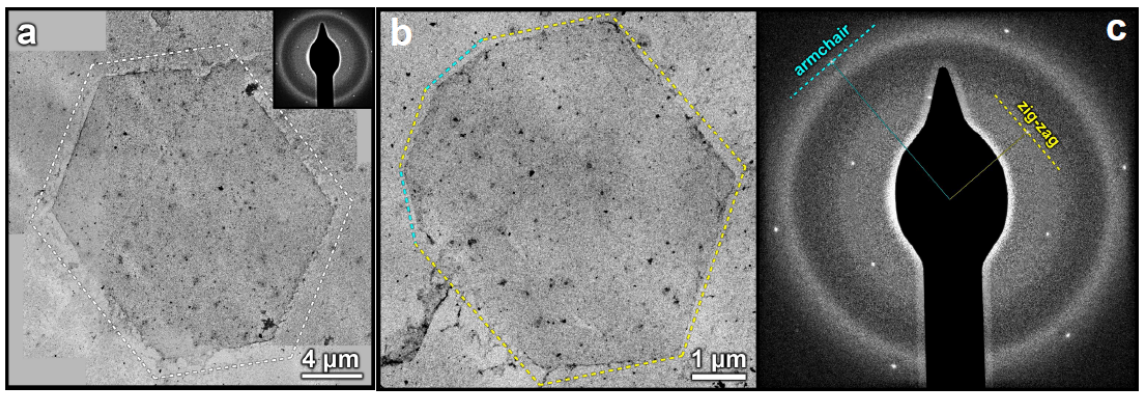

**Figure S5**. Additional examples of single-crystal graphene grains imaged by TEM/SAED. (a) Bright field TEM image and adjoining processed SAED pattern (inset) of a graphene grain. All the edges are nearly parallel to zigzag directions (white dashed lines). The line through the middle of the image is likely an artifact from graphene transfer with no effect on image interpretations. (b) Bright field TEM image and adjoining processed SAED pattern (c) of another grain. Most of its edges are aligned close to zigzag directions (yellow dashed lines). Two segments, however, are apparently parallel to armchair directions (cyan dashed lines) and are attributed to corners of zigzag edges that have evenly folded over.

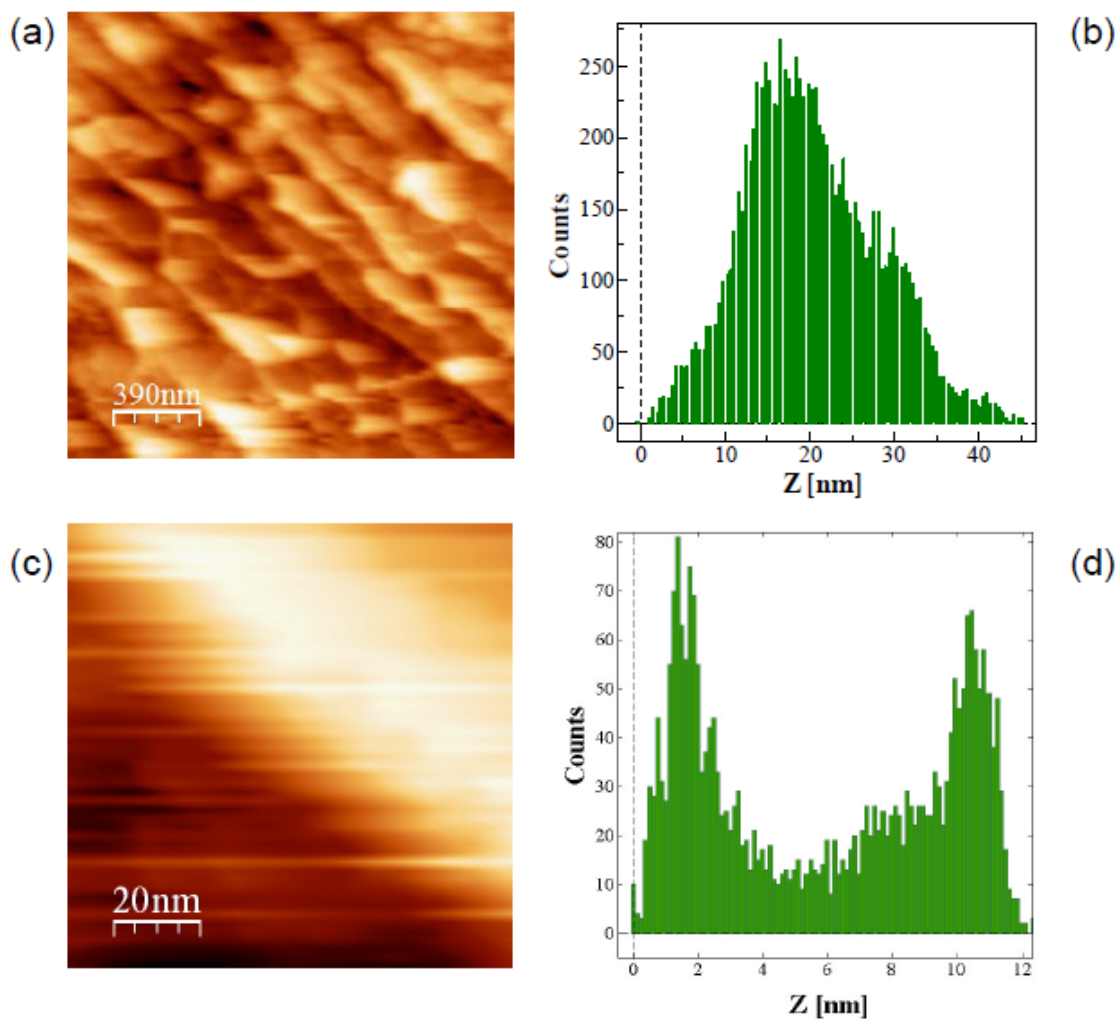

**Figure S6**. Surface height fluctuations of graphene synthesized on Cu, as characterized by scanning tunneling microscopy (STM). (a) An STM topography image of a 2 μm x 2 μm area in a graphene grain (in the vicinity of green square marked in main text Fig. 3). (b) Height histogram of all the pixels in image (a). The rms (root-mean-square) roughness is ~9 nm. Measurements from other areas of graphene yield similar roughness. A significant source of the observed roughness on graphene is attributed to the steps on the underlying Cu substrate, giving rise to the line texture seen in image (a). Such steps are nearly unobservable on Cu surface not covered (and protected) by graphene, due to the formation of a thin layer of native copper oxide. (c) A zoomed-in image from a smaller area in the same graphene grain shown in (a), containing only one Cu step. (d) Height histogram of the pixels in image (c).

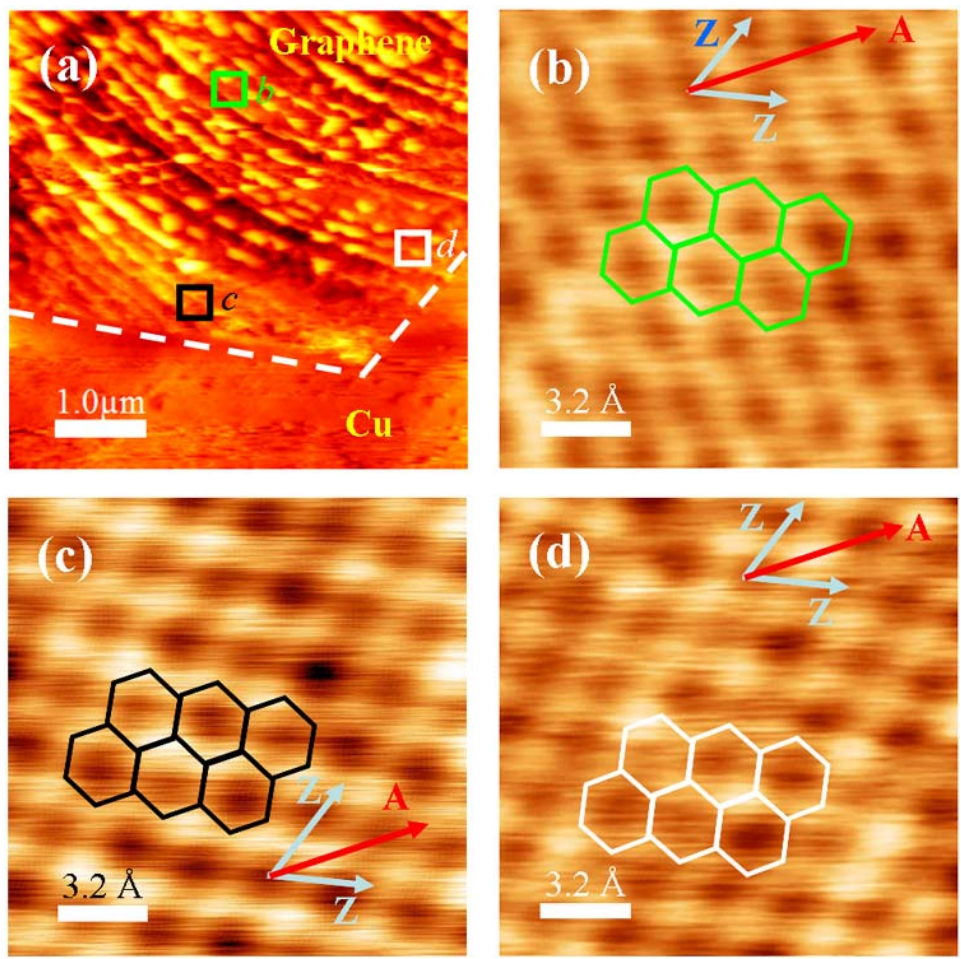

**Figure S7.** Scanning tunneling microscopy (STM) of a single-crystal graphene grain on Cu: (a) is the same image as main text Fig. 3a and (b-d) are unfiltered raw images corresponding to Fig. 3b-3d respectively.

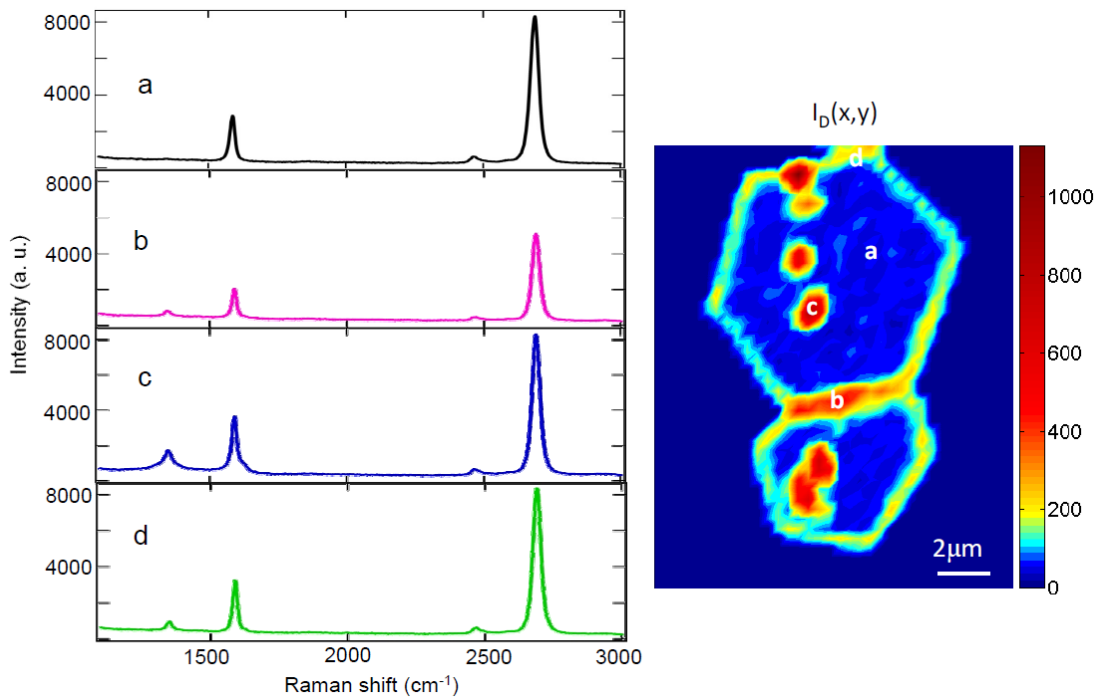

**Figure S8**. Individual Raman spectra taken from 4 representative locations from the graphene grain presented in main text Fig. 4d (reproduced on the right, with the 4 locations labeled): (a) inside a single-crystal graphene grain, (b) on the grain boundary between two coalesced graphene grains, (c) on a defect site, which could be a nucleation center, inside an individual grain, and (d) on the edge of graphene.

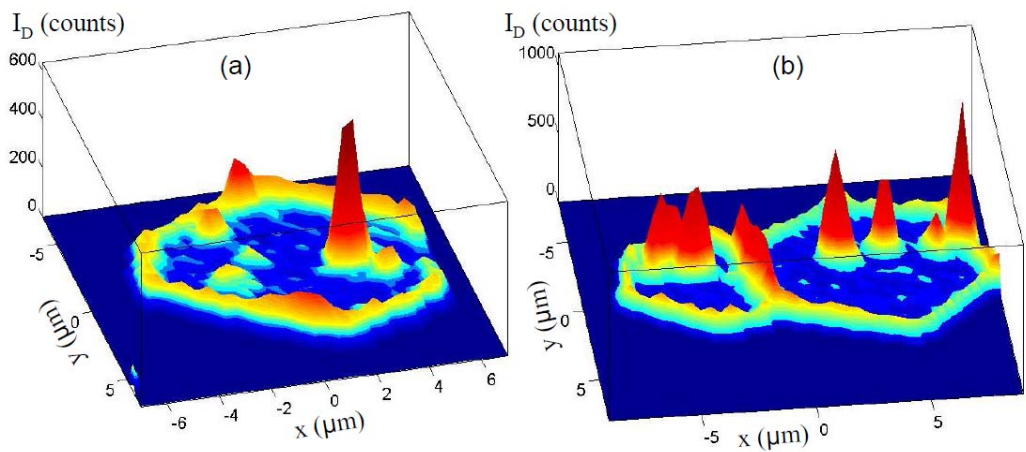

**Figure S9**. False-color 3D plots of the spatially dependent Raman "D" peak intensity ($I_D$) for the graphene grains studied in main text Fig. 4. Here (a) and (b) correspond to Figs. 4a and 4d respectively.

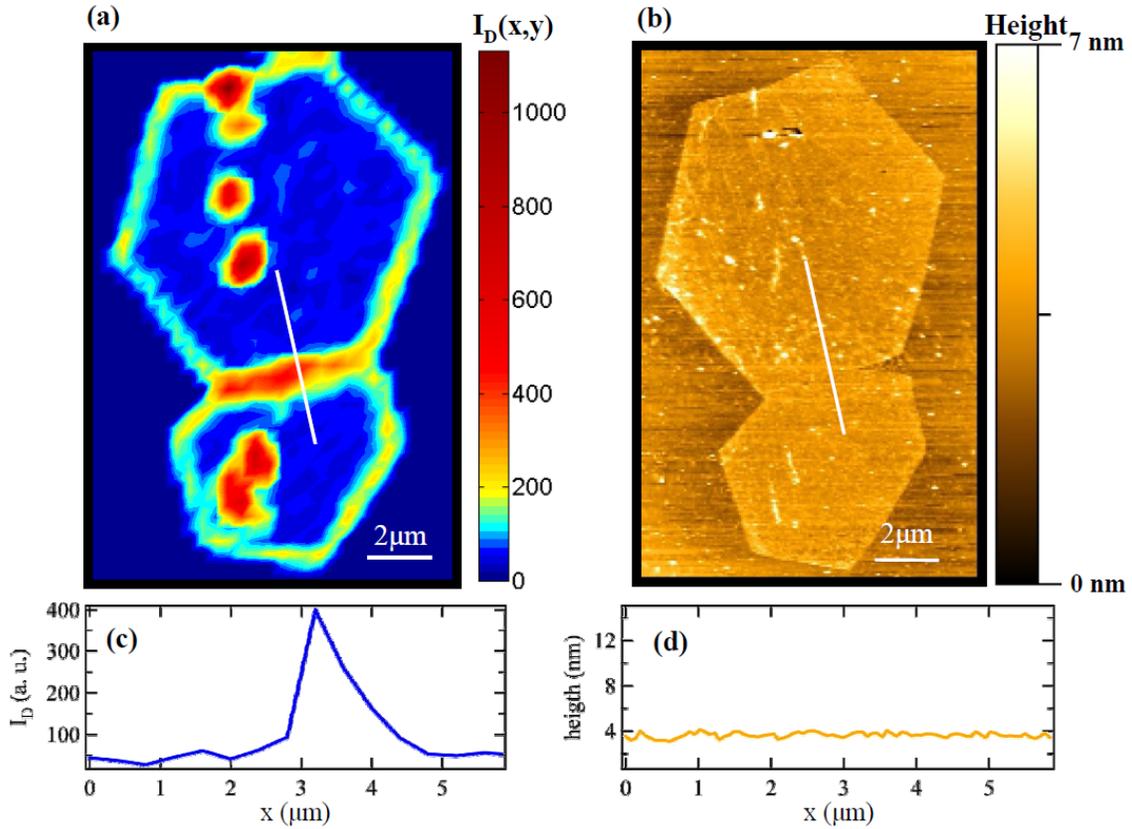

**Figure S10.** Comparing Raman "D" peak map (a) and atomic force microscopy (AFM) height image (b) of the same sample shown in main text Fig. 4(d). (c) and (d) are cross sectional profiles cut along the white lines in (a) and (b) respectively. The grain boundary does not give any observable features in AFM (b), but is clearly identified in Raman "D" map (a). AFM was carried out at ambient condition using the tapping mode.

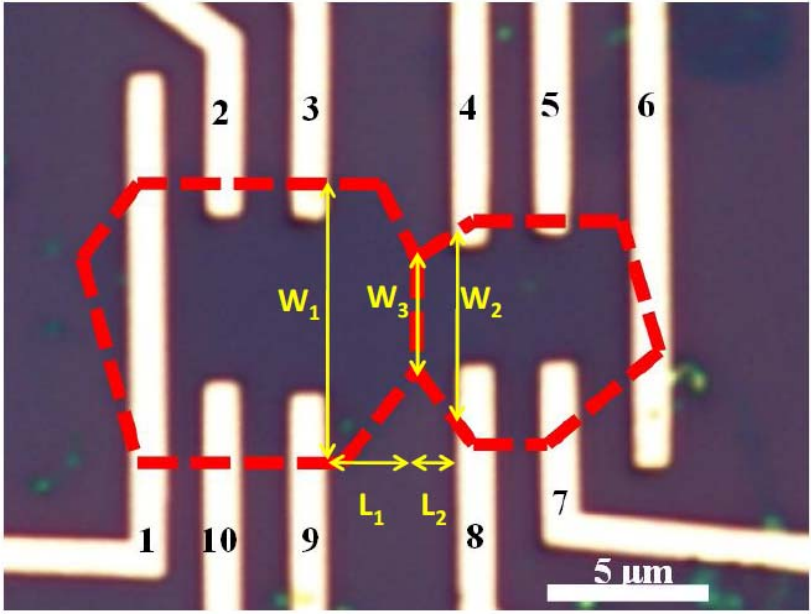

**Figure S11.** An illustrative example to show how the inter-grain effective resistivity ($\rho_{CG}$) and inter-grain series resistance (r) are defined. The optical image is that of the device shown in main Fig. 5. $\rho_{CG}$ is calculated from measured inter-grain resistance $R_{CG}$ (eg. between electrodes 8 and 9) as $\rho_{CG}=R_{CG}/(L/W_a)$. $W_a$ is the geometric mean of the sample width relevant for the inter-grain transport, calculated from $\frac{1}{W_a} = \frac{1}{L}\int_0^L \frac{dx}{W(x)}$, where $L=L_1+L_2$ and $W(x)$ is the sample width that varies along the direction (x) for inter-grain transport (here $W(0)=W_1$, $W(L_1)=W_3$, $W(L)=W_2$). The inter-grain series resistance (r) is calculated as $r = \rho_1 \int_0^{L_1} \frac{dx}{W(x)} + \rho_2 \int_{L_1}^{L} \frac{dx}{W(x)}$, where $\rho_1$ and $\rho_2$ are intra-grain resistivities of the left and right grain respectively, using values from intra-grain transport measurements. This r provides an estimate of what the inter-grain resistance would be if there were no grain boundary.

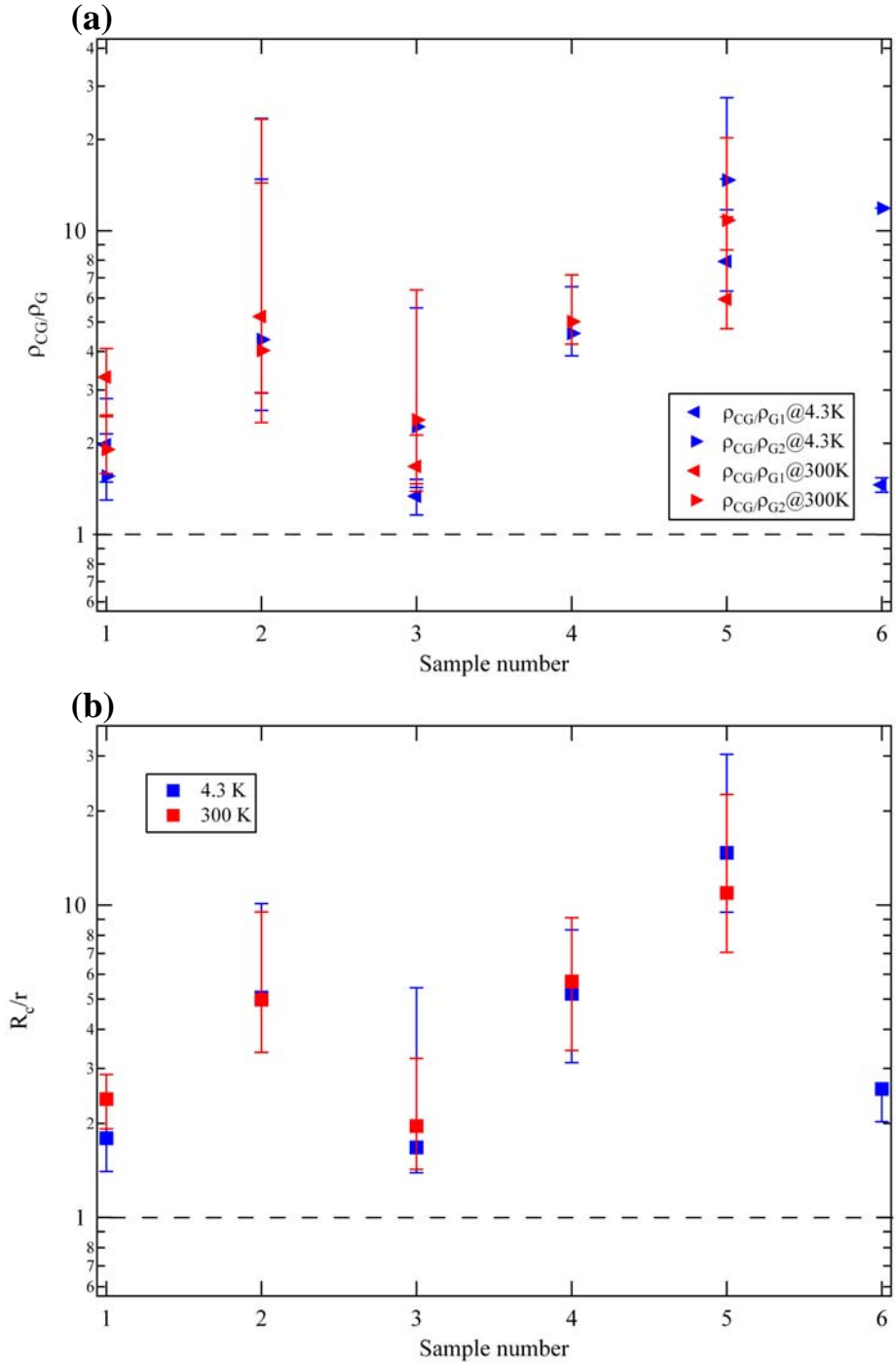

**Figure S12.** (a) Ratio ($\rho_{CG}/\rho_G$) of effective inter-grain resistivity ($\rho_{CG}$) to intra-grain resistivity ($\rho_G$, where G1 and G2 refer to the two graphene grains on each side of the grain boundary) measured in 6 devices. $\rho_{CG}$ is calculated from the measured inter-grain resistance ($R_C$) using the mean sample width ($W_a$) as defined in Fig. S11. (b) Ratio ($R_c/r$) of measured inter-grain resistance ($R_c$) to calculated (neglecting the grain boundary)

inter-grain series resistance (r, see definition in Fig. S11). Error bars reflect multiple measurements and/or using different electrode configurations (when available). Despite the variation in the results, grain boundaries were observed to always impede the transport (enhance the resistance) in graphene.

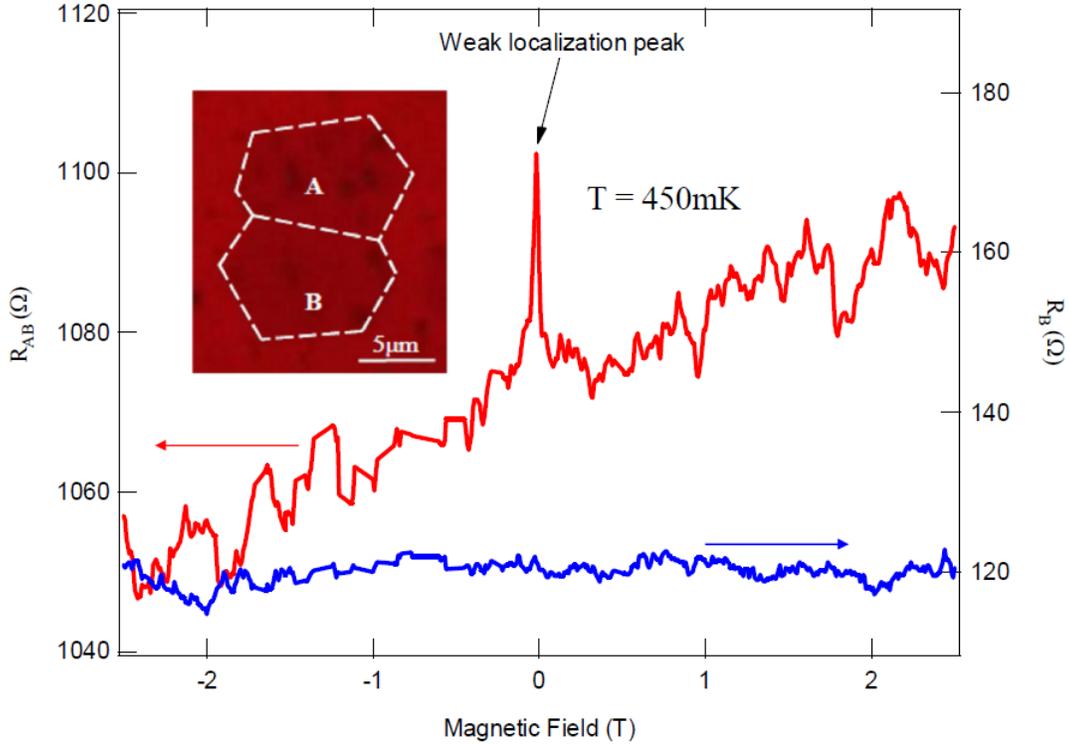

**Figure S13**. Four-terminal intra-grain and inter-grain magnetoresistances ($R_{xx}$) measured at 450 mK in a device ("b") consisting with two coalesced graphene grains with one grain boundary. Inset shows optical image of the sample before lithography. The inter-grain $R_{xx}$ ($R_{AB}$) displays a prominent weak localization (WL) peak, while no such WL feature is observed in the intra-grain $R_{xx}$ ($R_B$, measured within grain B). The raw data (without symmetrization) presented here contain a small Hall component (giving rise to a linear background, eg. in $R_{AB}$).

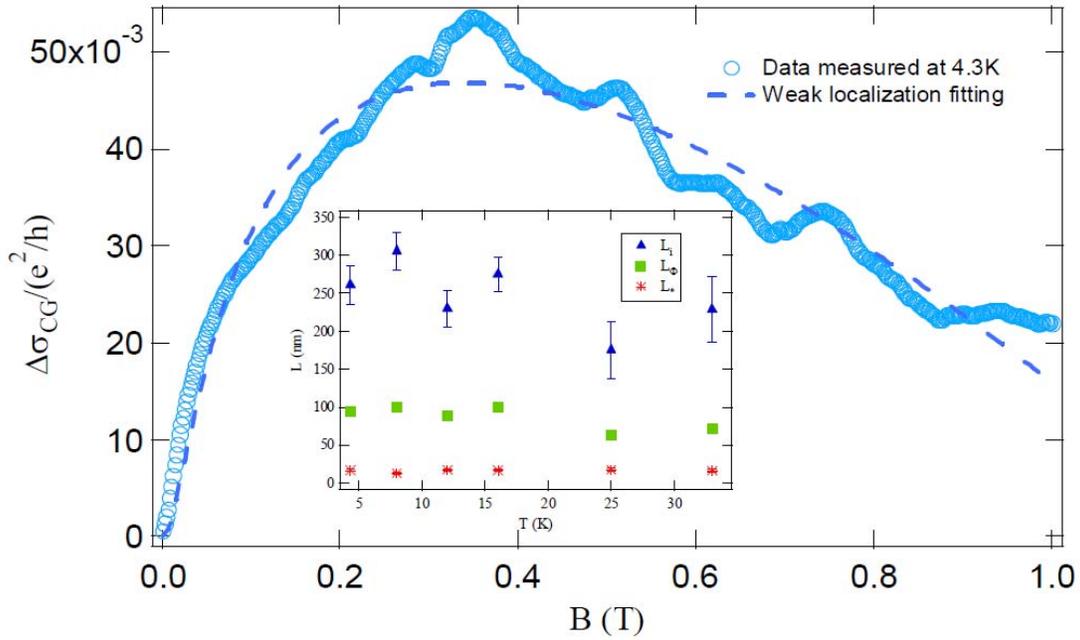

**Figure S14.** An example of weak localization fitting (dashed line) for cross-grain boundary (CG) magnetoconductivity ($\Delta\sigma_{CG}(B) = \sigma_{CG}(B) - \sigma_{CG}(B=0)$, normalized by $e^2/h$) measured for the device "a" shown in main text Fig. 5 at T=4.3 K. We fit the data to the equation describing weak localization in single layer graphene (Ref. S4)

$$\Delta\sigma(B) = \frac{e^2}{\pi h} \times \left[ F\left(\frac{B}{B_\varphi}\right) - F\left(\frac{B}{B_\varphi + 2B_i}\right) - 2F\left(\frac{B}{B_\varphi + B_i + B_*}\right) \right] \text{ where}$$

$$F(z) = \ln(z) + \Psi\left(\frac{1}{2} + \frac{1}{z}\right) \text{ and } B_{\varphi,i,*} = \left(\frac{h}{8\pi e}\right) L_{\varphi,i,*}^{-2} \text{ ($\Psi$ being the digamma function).}$$

Inset shows characteristic lengths (inelastic scattering/phase-breaking length $L_\varphi$, intervalley scattering length $L_i$ and intravalley scattering length $L_*$) extracted from such fittings for data measured at various temperatures.

**Supplemental references**:

[S1] X. Li *et al*., Large-Area Synthesis of High-Quality and Uniform Graphene Films on Copper Foils. Science **324**, 1312 (2009)

[S2] L. Colombo *et al*., Growth kinetics and defects of CVD graphene on Cu, ECS Trans., **28** (5) 109-114 (2010)

[S3] X. Li *et al*., Graphene Films with Large Domain Size by a Two-Step Chemical

Vapor Deposition Process, Nano Lett., 10, 4328-4334 (2010)

[S4] E. McCann *et al*, Weak-localization magnetoresistance and valley symmetry in graphene. Phys. Rev. Lett. **97**, 146805 (2006)

**Additional Acknowledgment**: The low temperature magnetotransport data in Fig. S13 were measured in a He-3 superconducting magnet system in National High Magnetic Field Laboratory (supported by NSF, DOE and the State of Florida). We thank J. Park, G. Jones and E. Palm for experimental assistance.